\documentclass[final,5p,times,twocolumn,authoryear]{elsarticle}
\usepackage{glossary}
\usepackage{config}
\usepackage{amssymb}
\usepackage{amsmath}

\journal{Visual Informatics}

\begin{document}

\begin{frontmatter}



\cortext[cor1]{Sicheng Song is the corresponding author.}

\title{SupplyNet: Supporting Visual Exploratory Learning in Supply Chain via Contextual Multi-Agent Simulation} 


\fntext[label2]{These authors contributed equally.}

\author[hkust]{Yanjia Li\fnref{label2}} 
\ead{ylifs@connect.ust.hk} 
\author[hkust]{Kelcy Kexin Han\fnref{label2}} 
\ead{khanac@connect.ust.hk} 
\author[cityu]{Tianrui Hu} 
\ead{tianruhu-c@my.cityu.edu.hk} 
\author[hkust]{Yi-Fan Cao} 
\ead{caoyifan@ust.hk} 
\author[hkust]{Huamin Qu} 
\ead{huamin@ust.hk} 
\author[hkust]{Sicheng Song\corref{cor1}} 
\ead{csescsong@ust.hk} 

\affiliation[hkust]{organization={The Hong Kong University of Science and Technology},
            city={Hong Kong},
            country={China}}
\affiliation[cityu]{organization={City University of Hong Kong},
            city={Hong Kong},
            country={China}}

\begin{abstract}
Simulation has long supported supply chain management instruction by letting learners observe network behavior and test decision strategies. Recent progress in LLM-driven agents opens new possibilities for richer, more adaptive simulations, but many existing systems still present abstract, opaque data that overwhelms learners and discourages active exploration. We introduce \textit{SupplyNet}, a gamified visual simulation system built on a contextual graph-based LLM multi-agent framework that models interdependent supply chain dynamics and provides responsive feedback through tiered challenges. \textit{SupplyNet} turns the simulation into a manipulable decision space by integrating an interactive network view of system state, a branching timeline for “what-if” exploration and comparison, and a task-oriented analysis console for structured performance breakdowns. Together, these visual components support counterfactual exploration, causal tracing, and comparative reasoning about outcomes. A user study suggests that \textit{SupplyNet} increases engagement and supports users’ perceived understanding of supply chain dynamics, highlighting the potential of pairing contextual multi-agent simulation with visualization to advance operational comprehension.
\end{abstract}


\end{frontmatter}


\section{Introduction}

\ac{scm} is a fundamental course in university curricula that equips students with uncertainty-aware decision making skills under the interdependency of supply chain dynamics~\cite{stadtler2014supply}. Yet it is difficult to learn from lectures alone: key concepts (e.g., disruption propagation, inventory oscillation, and trade-offs between service level and cost) emerge from the interaction of entities and time-varying constraints. Therefore, agent-based simulations have become a common pedagogical approach for \ac{scm}~\cite{swaminathan1998modeling, xue2005agent}, where business entities are represented as autonomous agents~\cite{swaminathan1998modeling}, allowing learners to experiment with decisions and observe dynamic network behaviors.~\cite{xue2005agent}. 

However, existing simulation systems are primarily designed for narrow concept demonstration, often overlooking realism and learner engagement. These systems typically represent complex networks~\cite{2023gvqa} as static diagrams~\cite{song2023vividgraph,song2024graphdecoder} and confine critical data to tables, creating a disconnect between abstract numbers and intuitive understanding~\cite{pfeffer2002end}. Such limitations are further compounded by the simplistic, rule-based agents, which fail to reflect the inner motivation of human beings in the real-world business context. Consequently, there is a lack of free space for students’ self-exploratory learning of supply chain knowledge, ultimately restricting the cultivation of their practical application and decision-making capabilities~\cite{pettigrew2016guest}.

To better understand these learning barriers, we conducted a formative study with 7 university students and 3 instructors, which informed the design of a contextual, feedback-rich, and comparison-driven exploratory simulation environment.

\revision{Based on these insights, we present \textit{SupplyNet}, a gamified visual simulation system designed for university students studying \ac{scm}, supporting them in exploring supply chain dynamics for engaging concept learning.}\ques{[R2-W1]} Powered by a \ac{llm}-based multi-agent framework, the system models the interdependent supply chain dynamics under customizable contexts and difficulty tiers. The system integrates gamified onboarding where users select roles and challenge levels, then define supply chain structures. During simulation, users can observe state changes, inject disruptions, make decisions, and receive responsive feedback.

\textit{SupplyNet} provides visual interfaces that render the environment as a manipulable decision space. The \textit{\textbf{Graph Structure View}} presents business entities and logistics as an intuitive network. For counterfactual reasoning, the \textit{\textbf{Timeline Tree Map}} tracks simulation progress and profitability and allows users to branch at any point to create parallel ``what-if'' scenarios for comparison. The \textit{\textbf{Analysis Console}} provides structured breakdowns of key performance metrics to reduce analytic burden. By combining contextual multi-agent simulation with decision-oriented visualizations, \textit{SupplyNet} supports learners in tracing causal propagation, testing strategies, and building operational understanding through sustained exploration.

We conducted a comprehensive user study to evaluate \textit{SupplyNet}. Participants reported improved learning support and a more engaging learning experience compared to a classic baseline simulation. We further observed various exploration styles and a reciprocal relationship between learning support and engagement, suggesting that contextual simulation and visual exploratory tools can reinforce one another in SCM learning.

Our contributions are threefold:
\begin{itemize}
    \item We characterize key barriers that limit exploratory learning in existing supply chain simulation tools, informed by a formative study with students and instructors.
    \item We propose \textit{SupplyNet}, a gamified visual simulation system for exploratory learning in supply chains, supported by contextual multi-agent simulation. \textit{SupplyNet} enables counterfactual exploration and causal reasoning through an interactive network view, a branching ``what-if'' timeline, and a task-oriented analysis console.
    \item We conduct a rigorous evaluation that demonstrates \textit{SupplyNet}’s effectiveness in enhancing learning support and engagement, outperforming the baseline.
\end{itemize}

\section{Related Work}
\subsection{Supply Chain Simulation}
Supply chain simulation has been indispensable for complex cost management, revealing downstream-upstream cascading effects and supporting decisions in inventory management, resource allocation and risk management~\cite{datta2011information, persson2002performance}.

\ac{abm} is widely adopted for supply chain simulation due to its strength in modeling entities' attributes and behaviors.~\cite{mohaddesi2022trust} models human-simulation interaction in supply chain shortage through statistical models to demonstrate how information sharing affects human decisions. Integrated with machine learning, ABM has greatly enhanced agents’ adaptability to dynamic environments. For example,~\cite{kotecha2024leveraging} and~\cite{liu2022multi} apply reinforcement learning in multi-agent systems to optimize inventory management amid ever-evolving market conditions. However, learning-based methods suffer from the challenge of acquiring large training datasets in practice~\cite{cannas2024artificial}. Furthermore, existing frameworks often neglect agent-user interactivity and agents’ decision interpretability, which are vital for users to better understand, trust, and influence agents’ decisions, thus promoting effective collaboration in dynamic supply chain contexts.

\subsection{LLM-based Multi-Agent Framework for Simulation}

\acp{llm} have emerged as a powerful technology for creating autonomous agents in multi-agent frameworks, featuring strong task performance, zero-shot adaptability to domain-specific context, interpretable reasoning processes, and natural language interaction with humans. In human-AI collaboration, \ac{llm}-based agents have been proven effective in boosting user productivity and understanding by facilitating tasks like writing~\cite{zhang2023visar, zhang2026adapt} or mitigating filter bubbles with diverse perspectives~\cite{zhang2024see}.

Building upon these foundational capabilities, researchers have increasingly applied \ac{llm} agents to simulate complex social and economic systems. Studies have shown that \ac{llm} agents can exhibit human-like behaviors in macroeconomic decision-making~\cite{li2024econagent} and visualization understandng~\cite{chen2025unmasking,song2026vizdefender} consistent with established principles, develop emergent competition dynamics aligned with sociological theories~\cite{zhao2024competeai, lin2024strategic, han2023guinea}, and demonstrate realistic cooperative behaviors in social and educational simulations~\cite{park2022social, gao2023s3,chen2026mis}. This body of work establishes that \ac{llm}-based agents are not merely task-solvers but can also be promising proxies for human actors in complex dynamic environments.

A key advantage of \ac{llm} agents for such simulations is their ability to adopt diverse personas and behavioral patterns via simple prompt engineering~\cite{choi2024picle, jiang2024personallm}. Their natural language outputs allow them to articulate their ``motivation'' and ``feelings'' in response to environmental stimuli, enhancing their human-likeness. For instance, \ac{llm} agents have displayed positive or concerned attitudes in simulated financial markets based on social media information, mirroring human emotional reactions~\cite{yuzhe2025twinmarket}. This interactivity improves simulation engagement and agent relatability, laying the groundwork for engaging simulated worlds.

The potential of \acp{llm} has also been extended to \ac{scm}. \cite{quan2024invagent} introduce an \ac{llm}-based multi-agent framework for a four-echelon chain-structured model, demonstrating the feasibility of \ac{scm} application with emergent dynamics and high flexibility. Our work builds upon this basic skeleton and further implements a graph-structured supply chain framework. We leverage the unique human-likeness, interpretability, and interactivity of \ac{llm} agents, and exploit their potential for gamified visual design, to develop an interactive tool that enhances learning engagement.


\subsection{Gamification in Educational Design}
Gamification applies game design elements in non-game contexts to increase engagement and motivate goal-oriented behavior. Prior work has demonstrated its effectiveness in improving learning efficacy~\cite{feger2019gamification}. By creating interactive and immersive learning experiences, gamification enriches system design and supports sustained user participation.

Existing studies have incorporated a variety of game elements into interactive systems to facilitate learning. For instance, contextualization frames learning process as a narrative-driven experiences to intensify immersion and engagement. In~\cite{du2024careersim}, users assume roles to explore different life trajectories within a simulated world. Reward mechanisms, including \textit{points, badges, and unlockables}, are frequently used to reinforce progress and encourage consistent task completion; in~\cite{kirchner2024outplay}, such mechanisms help users overcome procrastination. Social elements, such as \textit{competition} and \textit{leaderboards}, further promote active exploration through connections as demonstrated in~\cite{schade2023mapuncover}.

Gamification also has strong potential for learning \ac{scm}, where it facilitates active participation in simulations under uncertainty. Yet engagement and interaction design have received limited attention in this domain. \ac{scm} involves numerous quantitative factors, such as inventory levels, costs, prices, and delivery times—alongside abstract representations of relationships and performance~\cite{stadtler2014supply}. Current \ac{scm} simulation systems often represent these variables as raw numerical values in static tables and charts. A typical example is the Beer Game simulation~\cite{HBSPSimulationV3_online}, which presents order, shipment, and cost information in lengthy, time-ordered tables. This challenges users to track business performance and grasp the underlying supply chain dynamics, thereby narrowing usability and analytical insights. Although some systems, such as~\cite{skilldynamics2023beer}, offer more visually expressive interfaces, limited interactivity still keeps learners in a largely observational role, making it difficult to test hypotheses, revise decisions, and connect actions to downstream consequences through iterative exploration.

\textit{SupplyNet} addresses the gap by integrating game elements into the design, providing an engaging and responding learning environment for users to better perceive dynamics and refine decision-making strategies through repeated interaction.

\section{Formative Study}

We conducted a formative study to identify limitations in existing supply chain simulation systems and derive design considerations for improved engagement, usability, and interaction. Our study investigated: (1) challenges faced by students when learning \ac{scm}; (2) how current simulation systems support this process; and (3) desired features of an ideal simulation system.

\subsection{Participants}
We recruited 7 students (4 female, 3 male; S1-S7), including 3 undergraduates and 4 graduate students majoring in Accounting (S1, S3), Finance (S2), Marketing (S4, S5), and Operations Management (S6, S7). All participants had completed at least one supply-chain-related course and had relevant foundational knowledge. Students were compensated at \$20/hour. Additionally, 3 instructors (2 female, 1 male; E1-E3) with extensive cross-departmental experience in teaching supply-chain-related courses were invited from publicly accessible faculty directories at top-tier universities. Instructor details are shown in~\cref{tab:instructor_demographic}. They were compensated at \$50/hour.
\begin{table}[htbp]
    \centering
    \resizebox{\columnwidth}{!}{%
        \begin{tabular}{c|c|c|c|c}
        \hline
        Participant ID & Academic Title & Department & Experience (Years) & Location \\ \hline
        E1 & Associate Professor & Operations Management & 10+ & Hong Kong\\
        E2 & Assistant Professor & Marketing & 3 & United States\\
        E3 & Associate Professor & Industrial Engineering & 7 & Hong Kong\\ \hline
        \end{tabular}
    }
    \caption{Demographic and professional background of instructor participants.}
    \label{tab:instructor_demographic}
\end{table}
\subsection{Procedure}
The study followed standard ethical guidelines. Participants were informed of the study purpose, procedures, and their right to withdraw at any time. Informed consent was obtained prior to participation. All data were anonymized and used solely for research purposes. Each participant completed a one-hour semi-structured interview, conducted either in person or via Zoom. Sessions were audio-recorded with permission, transcribed, and stored with restricted access.
The interview followed a predefined outline with three phases. First, we collected background information on participants’ prior exposure to supply chain concepts and simulation systems. Students reflected on course experiences and past use of simulations, including perceived value and limitations, while instructors shared their professional background and observations of common challenges. Second, participants interacted with the ``Root Beer Game'', a single-chain simulation that illustrates the bullwhip effect and remains widely used~\cite{HBSPSimulationV3_online}. Each participant was given 15 minutes to explore the system in the role of a retailer, after which they assessed usability, interaction experience, and overall value. Finally, participants discussed expectations for an ideal supply chain simulation system, focusing on desired functionality, feedback mechanisms, and visual representations. 

\subsection{Findings}
Despite acknowledging the utility of simulations in illustrating core \ac{scm} fundamentals, participants prominently reported five key challenges that impede students’ learning and exploration, which we summarized as follows:
\subsubsection*{C1: Limited Contextual Cues Lead to Disengagement.}  Students consistently reported that the rigid, linear structure and random fluctuation of current simulations quickly create a disconnect from real-world complexity, leading to disengagement and boredom once the basic concepts are mastered (S3, S4, S6). S3 stated, \textit{``I would totally lose the interest to keep playing after mastering the basic concepts and demand patterns.''} This disconnect is lessened when the environment is contextualized. S4 echoed this, \textit{``I would have more fun when playing if the environmental changes are contextualized. For example, the demand fluctuation is subject to seasonal change or quality concerns.''} Instructors (E1-E3) confirmed that the lack of an engaging narrative or evolving complexity undermines the goal of long-term skill reinforcement (E1: \textit{``As a simulation game, it is important to tell a fun and interesting story to keep students engaged.''}). Advanced learners also expressed a desire for systems that could support new and challenging content to maintain engagement over time (S1, S2, S5, S6). As S2 stated, \textit{``It would be much more interesting if some recent events in life can be added to the simulated environment.''}. 
\subsubsection*{C2: Insufficiency of Timely and Actionable Feedback.}
A major source of student frustration was the absence of timely and specific feedback on their in-game decisions, hindering the learning process. Students felt "lost" and were often unaware of their performance mid-game, which led to random decision-making as the simulation progressed (S3: \textit{``I feel lost in the middle of the game for not knowing how good I perform.''} S4: \textit{``I received little hints on whether I have been making good or bad decisions in the game.''} ). While final summary statistics on earnings were available, most participants (S2-S6) found these numbers insufficient for evaluating their overall business management performance. Instructors acknowledged this critical gap, noting the impossibility of providing detailed, personalized feedback to every student due to high labor and time costs (E2: \textit{``However, it is almost impossible for us to provide detailed feedback to everyone due to high labor and time costs.''}). Consequently, students felt little incentive to replay the simulation with an objective of self-improvement.

\subsubsection*{C3: Constraint of Alternative Decision Exploration.}
All student participants (S1-S7) expressed a strong, inherent curiosity to explore how alternative planning decisions would lead to different consequences, but they felt constrained by the restricted operative space of existing systems. A primary limitation was the difficulty of comparing outcomes between different playthroughs. S5 commented that \textit{``Although I could always restart the simulation, it is challenging for me to compare the outcomes between different decision-making without a dedicated feature to place different timelines side-by-side for direct analysis.''} This finding aligns directly with instructors' teaching objectives. E2 confirmed that explicit support for varying one or two key factors (\eg, price or order quantity) and observing the differential outcomes is a valuable component of the counterfactual analysis prized in business studies (E2: \textit{``It contributes to the counterfactual analysis we valued in school.''}).

\subsubsection*{C4: Opaque Agent Reasoning and Interdependencies.}
Student learning was hampered by the opacity of the supply chain network’s internal dynamics. Simply observing evolving numerical factors was deemed insufficient for underlying understanding (S1, S3). Participants desired insight into the mutual effects within the simulation and the rationale driving computer-controlled agents' behavior (S3: \textit{``I am not only curious about how decisions affect the environment, but also how business entities are motivated to react to the environmental changes.''} ). S6 specifically wished for agents to \textit{``think aloud,''} elaborating that visibility into an agent's mind makes the simulation more realistic and helps grasp a manager's thinking. Instructor E2 highlighted that making agent reasoning transparent serves a dual educational purpose: it helps students \textit{``understand the inter-dependent nature of supply chain networks''} and the explanations themselves \textit{``serve as a helpful reference for students to manage their own business.''}

\subsubsection*{C5: Inadequate Visualizations for Analysis.}
The visual design and presentation of information were a consistent concern (S1-S7, E1-E2). Instructors noted that supply chain environments are information-dense, posing considerable challenges for students’ information processing (E2). S6 shared that previous simulations suffered from \textit{``messy data organization, which brought me much trouble to find environmental changes and check business performance when playing.''} Beyond basic data organization, most students (S3-S7) and instructors (E1-E2) expressed a strong need for task-oriented data visualization, including time-series and correlation analyses. S3 explicitly requested, \textit{``I wish to observe the evolution of key metrics over time, not only to understand the short- and long-term effects of decisions but also to compare the correlation between factors.''} E1 noted that intuitive visualization is essential to facilitate the financial and operational analyses that instructors require students to conduct for evaluating management efficiency.

These findings highlight the evolving learning needs associated with supply chain simulations and underscore the significance of more advanced visual-dynamic graph agent simulations designed to support exploratory learning.
\subsection{Design Goals}

Drawing upon the challenges identified in the formative study (C1--C5), we formulated three high-level design goals to bridge the gap between abstract supply chain concepts and practical experimental learning, guiding our subsequent system implementation. These goals articulate the necessary interaction strategies and visual capabilities, directly informing the subsequent system design.

\subsubsection*{DG1: Provide Immersive Context and Immediate Feedback.}
Our formative study revealed that static, numerical simulations detach students from the urgency of real-world supply chains (C1), while delayed feedback hinders their ability to evaluate decisions (C2). An effective learning environment would ground decision making in realistic operational contexts and evolving challenges rather than isolated numerical updates. Such context could help learners remain engaged beyond initial concept acquisition and would encourage iterative experimentation. The system should also provide timely and interpretable feedback so that learners could quickly connect actions and external disruptions to downstream consequences, adjust strategies, and maintain a sense of progress.
\subsubsection*{DG2: Support Counterfactual Exploration and Comparative Analysis.}
Participants reported frustration with the linear progression of traditional simulations, which prevents them from revisiting past decisions or directly comparing outcomes (C3). Therefore, we aim to lower the cost of experimentation and facilitate ``what-if'' analysis through iterative trial-and-error learning. The design should not only enable learners to test hypotheses over time (e.g., by changing ordering policies or responding to disruptions), but also create a structured way to compare decision paths and make alternative outcomes easy to notice and reflect on. This would support counterfactual analysis and the development of transferable decision heuristics.
\subsubsection*{DG3: Intuitively Visualize Network Structure and Dynamics.}
Given the multi-echelon dependencies (C5) and opaque interactions between entities (C4) in information-dense supply chains, traditional tabular interfaces often cause cognitive overload. To assist students in forming accurate mental models, the interface would present network structure and evolving system state in a clean and intuitive manner. It would make flows of materials and information between entities visually traceable, helping learners grasp the overall state at a glance and follow how changes propagate through the network. Furthermore, computer-controlled business managers would be assigned human-like personas that can vividly articulate their reasoning and motivations, making interactions more authentic and insightful. Learners should also benefit from data analytic support for trend inspection and structured comparison across alternative operations, helping them reason about correlations among factors and relate decisions to long-term impacts.


\begin{figure*}[ht]
  \centering
  \vspace{-10pt}
  \includegraphics[width=\textwidth]{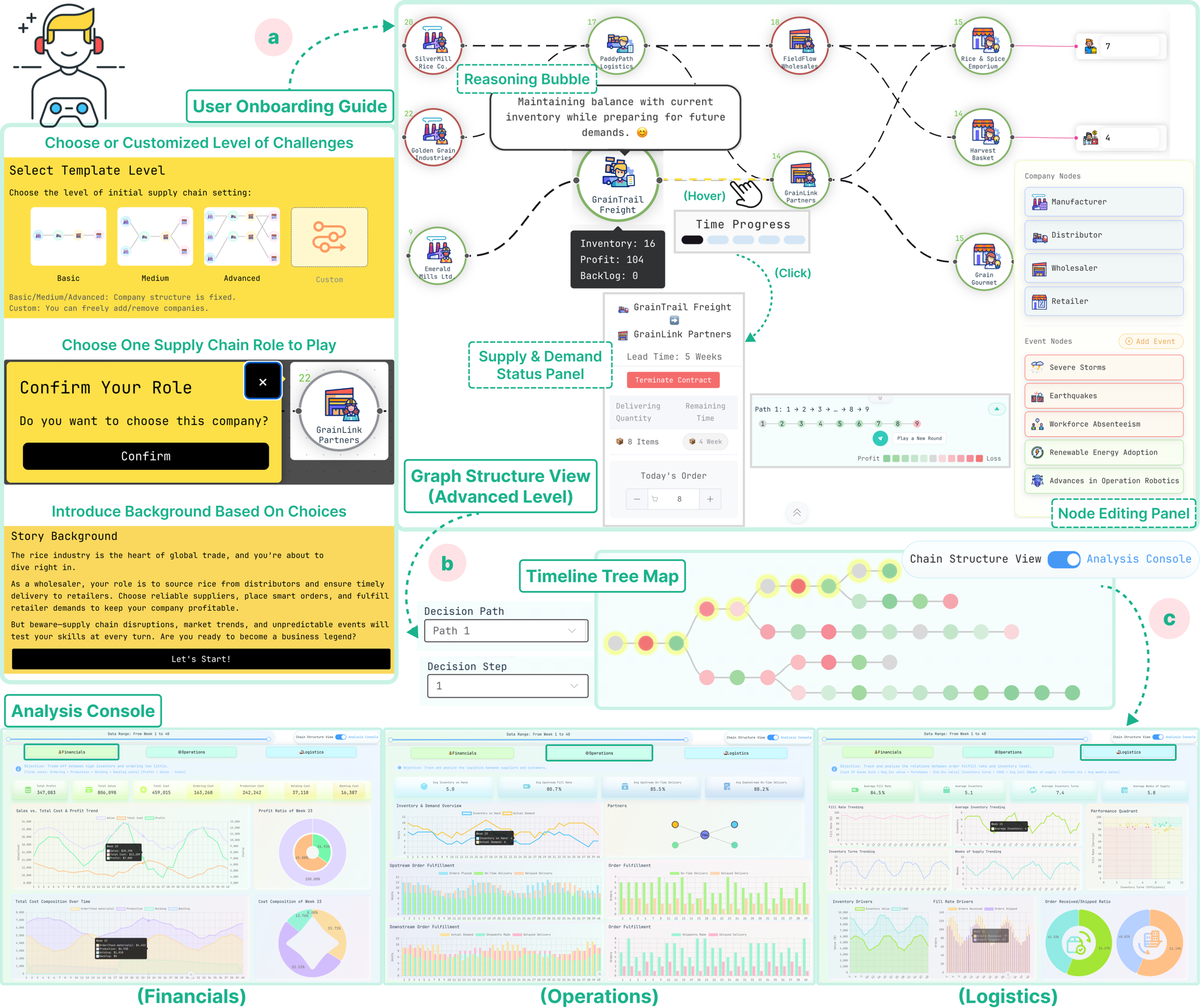} 
  \vspace{-10pt}
  \caption{\revision{Overview of the \textit{SupplyNet} interface. 
(1)~Users begin with the onboarding guide, selecting a challenge level and company role before entering the \textit{Graph Structure View}. 
(2)~The \textit{Timeline Tree Map} records decision paths and supports branching for ``what-if'' exploration. 
(3)~Users switch to the \textit{Analysis Console}, which provides three dashboards: 
(a)~financials, (b)~operations, and (c)~logistics, 
for inspecting performance over time and comparing outcomes across timelines.}\ques{[R2-M1]}}
  \label{fig:system_overview}
  \vspace{-10pt}
\end{figure*}

\section{System Overview}
\subsection{System Pipeline}
\textit{SupplyNet}'s learning objective centers on inventory management, guiding students to understand dynamic supply chain interdependencies by making sequential decisions on order placement and supplier selection and observing the resulting trade-offs in multiple factors.

\begin{figure*}[ht]
  \centering
  \vspace{-10pt}
  \includegraphics[width=\textwidth]{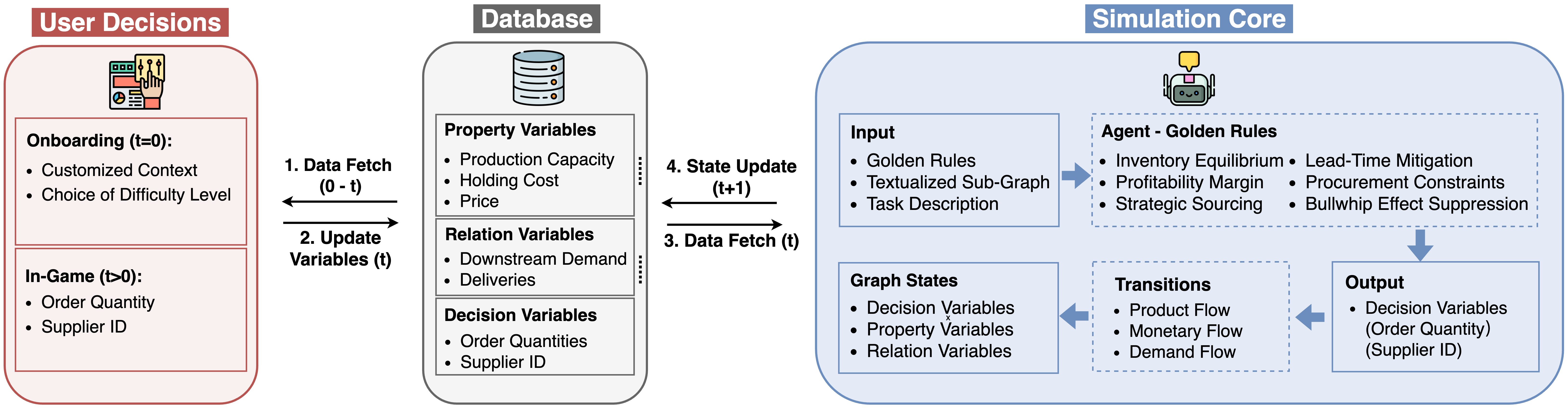}
  \vspace{-10pt}
  \caption{\revision{The bi-directional data flow in \textit{SupplyNet}.}\ques{[R1-Q4]}}
  \label{fig:system_architecture}
  \vspace{-10pt}
\end{figure*}

\revision{The pipeline follows an explicit bidirectional loop between an \ac{llm}-based multi-agent simulation core and an interactive visual interface. The simulation core(\ie, the multi-agent framework) maintains the evolving supply chain state as a directed graph: nodes encode entity attributes (inventory, price, and production capacity), and edges encode supply relationships and logistics (lead times, downstream demand, and arriving deliveries).}\ques{[R1-Q2, R1-Q4]} In each period, agents make sequential decisions on order placement and supplier selection; they are queried from downstream to upstream (retailers $\rightarrow$ wholesalers $\rightarrow$ distributors $\rightarrow$ manufacturers) to reflect demand propagation. \revision{To keep agent prompting scalable and context-aware, the simulation core performs agent-specific subgraph retrieval by extracting a localized context around each entity, including direct partners, relevant attributes, and active events. The retrieved subgraph is then textualized and combined with heuristic ``Golden Rules'' as agent input, enabling agents to produce role-specific actions along with brief motivation explanations. The agents return structured outputs containing decision variables (\eg, order quantities, supplier IDs). Together with the updated system states, it is saved as temporal snapshot to the database. The front-end reads data from the database to present real-time visualization on the user interface. Conversely, user decisions made on the interface are sent back to the database as an update of the system states. It is fetched by the simulation core when querying agents for decision-making. The overall data flow is illustrated in~\cref{fig:system_architecture}.}\ques{[R1-Q4]}\revisionn{At a more abstract level, this pipeline can be viewed as a reusable simulation architecture for dynamic relational learning environments. The graph represents domain entities, dependencies, and time-varying states; agents and learners observe relevant subgraphs, take domain-specific actions, and trigger state transitions; and the database stores temporal snapshots that support replay, branching, and comparison. In SupplyNet, this architecture is instantiated with business entities, logistics links, inventory-related states, ordering decisions, disruption events, and operational KPIs. Other subjects can reuse the same architecture by replacing the domain schema, transition rules, agent heuristics, event types, and learning metrics. We return to this abstraction in~\cref{sec:genralizavbility} to discuss cross-subject instantiations.}\quess{Q1}

\subsection{Workflow and Coordinated Views}
\textit{SupplyNet}'s interface is organized as a workflow across four views (Fig.~\ref{fig:system_overview}). Users begin with user onboarding, where they select a company role and challenge level and receive a narrative context, grounding subsequent decisions in a realistic setting (\textbf{DG1}). \revision{They then manage the simulation in the \textit{Graph Structure View}, which serves as the primary sandbox for monitoring network status, inspecting entities and links, and executing decisions (Fig.~\ref{fig:system_overview}(1)); the \textit{Node Editing Panel} supports constructing custom networks and injecting contextual events, while the \textit{Supply \& Demand Status Panel} provides link-level details for inspecting orders, deliveries, and fulfillment.}\ques{[R1-Q1]} After each step, updated states and lightweight outcome cues are surfaced to help learners connect actions and disruptions to downstream consequences (\textbf{DG1}), while maintaining interpretable network reasoning about propagation (\textbf{DG3}).

As the simulation progresses, actions and outcomes are recorded in the \textit{Timeline Tree Map}, which preserves decision paths and enables users to revisit earlier states and branch parallel trajectories for ``what-if'' exploration and comparison (Fig.~\ref{fig:system_overview}(2)), enabling counterfactual reasoning (\textbf{DG2}). Users can switch to the \textit{Analysis Console} to examine structured dashboards of key performance indicators over time and compare outcomes across timelines (Fig.~\ref{fig:system_overview}(3)), supporting diagnosis and reflection on longer-term impacts (\textbf{DG2}, \textbf{DG3}). Learners iterate by returning to the \textit{Graph Structure View} to revise decisions or introduce disruptions based on insights from the \textit{Timeline Tree Map} and the \textit{Analysis Console}.

\section{Visualization and Interaction Design}
In this section, we first introduce the common network visual substrate that makes supply chain structure and dynamics legible, and then present the contextualization and analysis tools that build upon it (Fig.~\ref{fig:system_overview}).

\subsection{Graph Structure Workspace}
To support learners in reasoning about complex dependencies and interdependent dynamics (DG3), \textit{SupplyNet} centers interaction around the \textit{Graph Structure View}. Inspired by the ``Visual Information Seeking Mantra''~\cite{shneiderman2003eyes}, the view prioritizes network overview and supports progressive disclosure of details through interaction. This shifts learners from isolated tables to a topological representation that makes structure, propagation, and local constraints simultaneously inspectable.

The \textit{Graph Structure View} serves as the primary workspace for operating the simulated supply chain. It visualizes the environment as a directed graph, where nodes represent business entities across stages and edges represent logistical relationships. To reduce cognitive overload in information-dense environments, the view adopts lightweight visual status encodings~\cite{song2024gvvst} that provide immediate summaries while preserving access to detail. Nodes are color-coded to reflect real-time operational outcomes: green indicates profitability, red represents losses due to backlog or other operational issues, and gray signals a shutdown state (Fig.~\ref{fig:visual_encoding}(a)). \revision{These status encodings also function as lightweight visual reward cues~\cite{feger2019gamification}, giving learners an immediate sense of progress or concern without requiring them to open detailed panels.}\ques{[R2-W5]} Edges visually distinguish active versus planned relationships and animate to indicate transportation progress: active orders are shown with solid lines, and planned partnerships with dashed lines. When hovering over an edge, a compact progress indicator appears to support rapid scanning without opening additional panels (Fig.~\ref{fig:visual_encoding}(b)).

The view supports progressive disclosure through details-on-demand, allowing users to scan the simulation at multiple levels of granularity. Selecting a node reveals its local operational state, such as inventory, capacity, backlog-related costs, and other decision-relevant attributes. Selecting an edge opens the \textit{Supply \& Demand Status Panel}, which focuses on the relationship between two connected entities by exposing lead time and fulfillment progress. Together, these interactions allow learners to quickly locate bottlenecks and follow how local constraints propagate through upstream and downstream dependencies.

To support scenario construction and controlled interventions, the \textit{Node Editing Panel} enables users to modify network structure and attach contextual disruptions directly onto entities. Users may add entities to construct alternative hierarchies, or inject events by dragging event icons onto nodes. Injected events are visually highlighted on the graph, helping users connect a local shock to subsequent network-wide changes as the simulation evolves.

Finally, to mitigate the perceived opacity of computer-controlled managers, \textit{SupplyNet} exposes agent decision rationale through an on-demand \textit{Reasoning Bubble}. \revision{When users click an automated entity, the bubble appears above the entity and presents a brief textual explanation of the agent's motivation for its current action and a compact affective cue that reflects current business performance.}\ques{[R1-Q1]} Specifically, the motivation is shown as a concise sentence, while the affective state is mapped to an emoji to preserve interface cleanliness. This design keeps the primary workspace lightweight while still providing interpretable hooks for learners to connect agent behaviors to evolving system conditions.
\begin{figure}[hbt]
  \centering   
  \includegraphics[width=\linewidth]{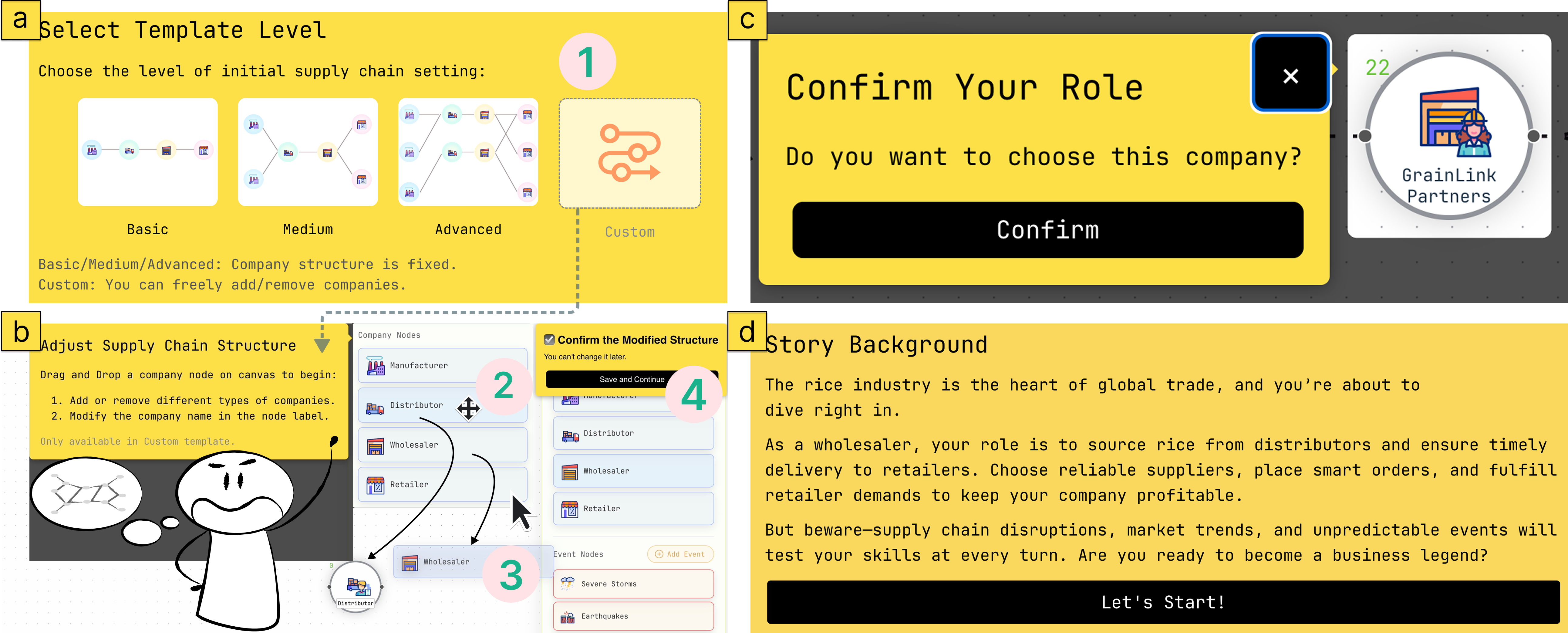}
  \vspace{-10pt}
  \caption{Onboarding Process. (a) Challenge selection. (b) Custom network assembly. (c) Role selection with narrative context.}
  \label{fig:onboarding}
  \vspace{-10pt}
\end{figure}
\subsection{Onboarding, Challenges, and Contextual Feedback}
To transform abstract operations into a situated learning experience (DG1), \textit{SupplyNet} integrates narrative framing, tiered challenges, and real-time feedback loops into the interaction flow. This design aims to sustain engagement beyond initial concept acquisition and help learners rapidly connect decisions and disruptions to downstream consequences.

\subsubsection*{Immersive Onboarding and Tiered Challenges.} 
Before the simulation begins, the \textit{User Onboarding Guide} (Fig.~\ref{fig:onboarding}) establishes context and calibrates complexity to learners' needs. During onboarding, users first choose a challenge level that matches their experience. The predefined levels progressively increase structural complexity, while a sandbox option supports free-form network assembly for open-ended exploration. A basic level introduces a linear four-stage chain to focus attention on core inventory concepts without overwhelming learners. Intermediate and advanced levels introduce a hub-and-spoke network and additional customization, encouraging strategic sourcing and resilience-oriented responses. Users then select a specific company role, and the system provides a short role-specific narrative that frames the simulation as a first-person managerial task with an explicit mission, making decision making feel purposeful and situated rather than purely numerical. \revision{This narrative contextualization is a core gamification strategy~\cite{dominguez2013gamifying} that transforms abstract operational tasks into situated, goal-directed activities.}\ques{[R2-W5]}
\begin{figure*}[htbp]
  \centering
  \includegraphics[width=\linewidth]{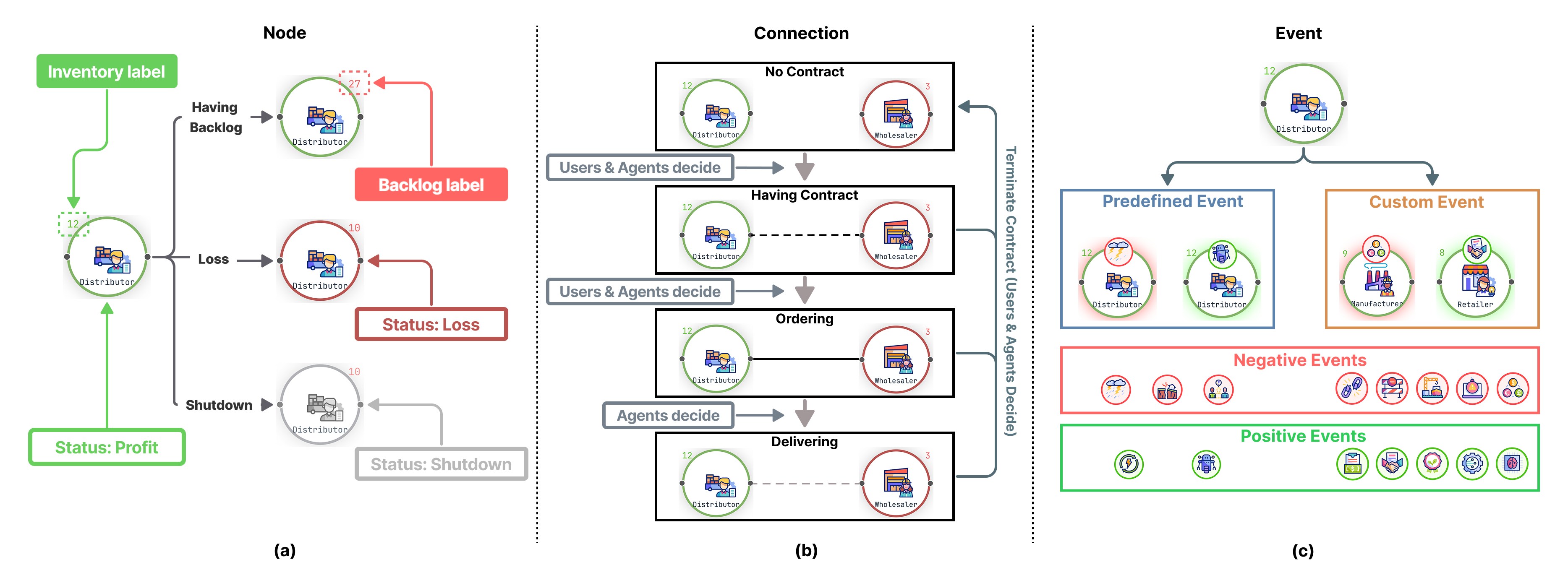}
  \vspace{-10pt}
  \caption{Visual Encodings. (a) Node status summary. (b) Link relationships and logistics progress. (c) Event attachment and impact visualization.}
  \label{fig:visual_encoding}
  \vspace{-10pt}
\end{figure*}
\subsubsection*{Events and Real-time Feedback.}
Contextual volatility is introduced through dynamic events. \revision{At the lower section of the \textit{Node Editing Panel} there is an event palette offering a set of predefined scenarios, such as natural disasters, workforce shortages, and technological upgrades, with positive or negative effects on factors such as inventory and logistics.}\ques{[R1-Q1]} Dragging event icons onto specific nodes simulates their effects, with immediate visual feedback highlighting positive impacts in green and negative impacts in red (Fig.~\ref{fig:visual_encoding}(c)). Furthermore, users can create their own disruptions by clicking the ``Add Event'' button. This allows them to define a custom event’s name, its specific impact, magnitude, and duration. This functionality helps users understand how disruptions propagate through the network while evaluating possible mitigation strategies.

\textit{SupplyNet} provides real-time feedback tightly coupled with each decision cycle. When a user confirms a decision for a given week, the backend advances the simulation and updates deliveries, inventory, and sales-related outcomes; key outcome indicators such as profit and backlog signals are refreshed immediately in the interface. When a user switches to the \textit{Analysis Console} to examine the history, the system retrieves historical data from the database and performs real-time calculations to generate statistical summaries. \revision{This immediate feedback loop mirrors the tight action–reward cycles characteristic of gamified learning environments, supporting iterative learning and creating an incentive mechanism by enabling learners to quickly evaluate decision quality and maintain a sense of progress in evolving contexts (DG1).}\ques{[R2-W5]}

\begin{figure}[hbt]
  \centering
  \includegraphics[width=\linewidth]{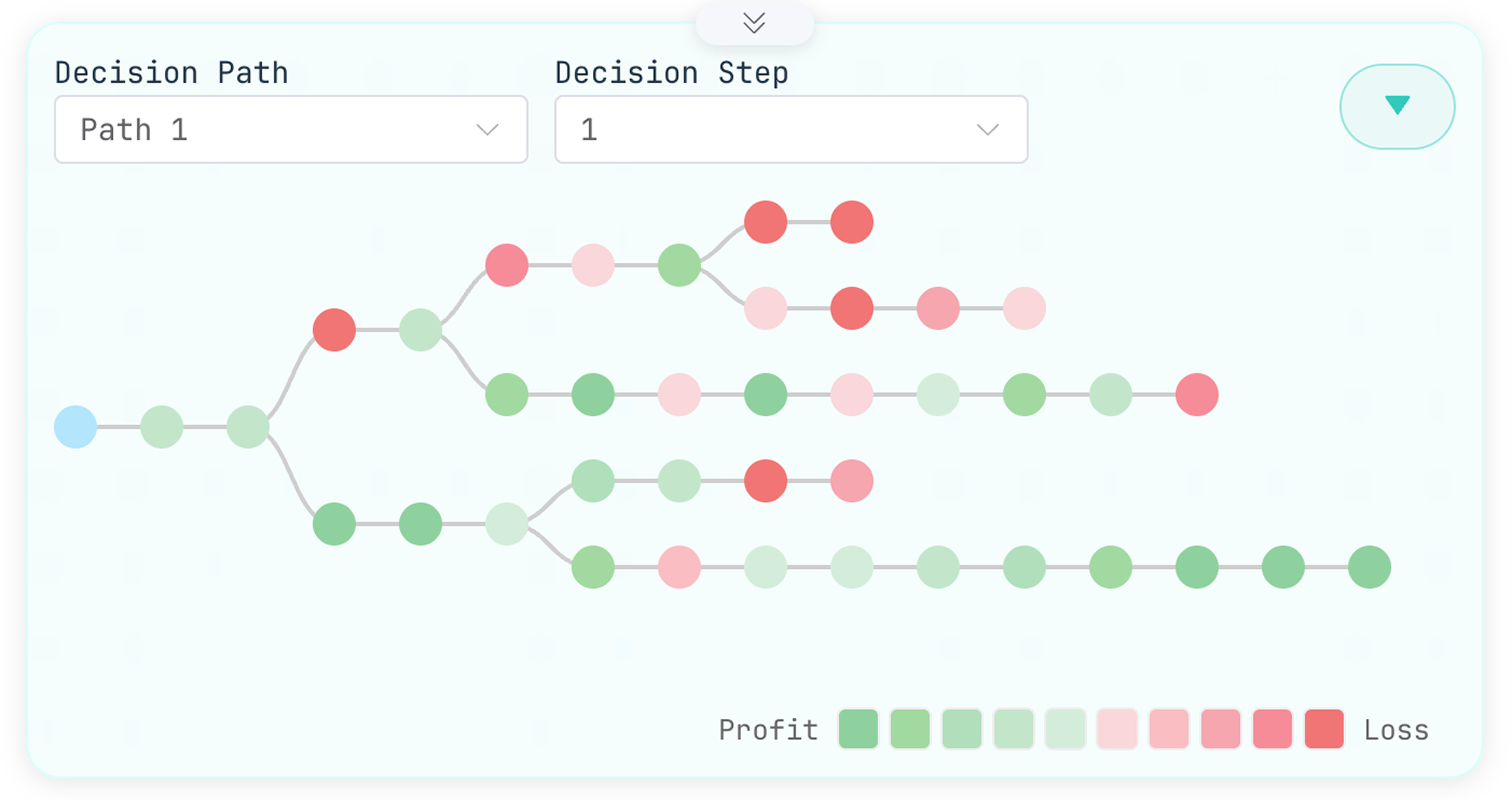}
  \vspace{-10pt}
  \caption{The \textit{Timeline Tree Map} visualizes the simulation history, enabling users to revert to past states and branch new timelines for comparison.}
  \label{fig:timeline}
  \vspace{-10pt}
\end{figure}
\subsection{Timeline Tree Map for Counterfactual Exploration}
A key contribution of \textit{SupplyNet} is enabling non-linear exploration of decision consequences. Drawing on counterfactual thinking~\cite{roese1997counterfactual} and external cognition~\cite{scaife1996external}, the interface externalizes simulation history so learners can compare ``what happened'' with plausible alternatives without relying on memory.

The \textit{Timeline Tree Map} (Fig.~\ref{fig:timeline}) records simulation progression as a branching structure rather than a linear log. Each step node represents a decision point, visually encoded with an outcome summary to support fast scanning of successful versus problematic periods. Users click any past step to revert the simulation state to that point, modify a decision, and resume the simulation to spawn a new parallel branch. When a past decision is modified, downstream steps on the affected branch are reset to preserve consistency, ensuring that subsequent outcomes reflect the revised history. \revision{This branching mechanism functions as an explicit ``save and branch'' interaction, echoing the save and reload mechanic familiar from digital games, lowering the cost of experimentation while maintaining provenance of alternative strategies.}\ques{[R2-W5]} By preserving parallel trajectories, the timeline makes comparison concrete and repeatable, assisting structured ``what-if'' exploration rather than ad hoc replay.

The \textit{Timeline Tree Map} is designed for both backtracking and comparison. Users can switch between branches to examine how alternative decisions shape subsequent outcomes. This design helps direct contrast across strategies and encourages users to articulate why outcomes diverge, strengthening transferable heuristics for inventory and fulfillment management (DG2).

\begin{figure*}[tbp!]
    \centering
    \includegraphics[width=\textwidth]{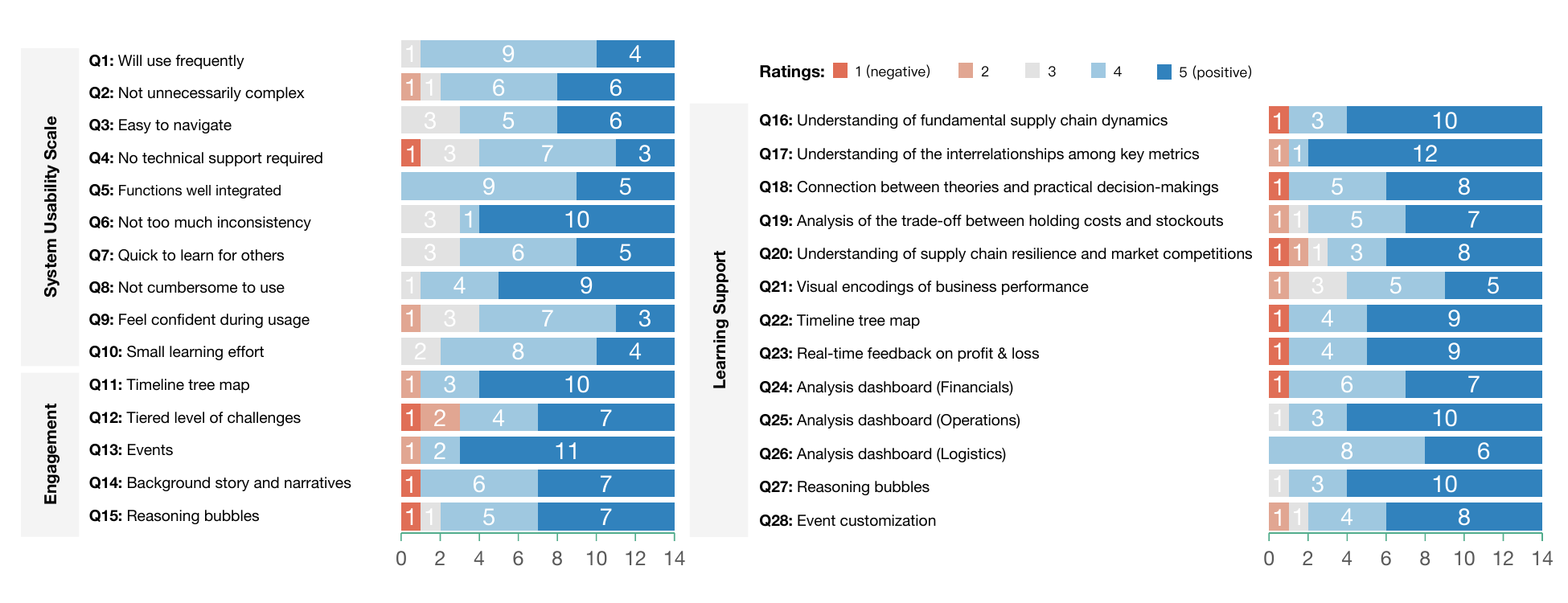}
    \vspace{-10px}
    \caption{Results of questionnaires for the system's evaluation metrics, assessing three key dimensions: the \ac{sus} (Q1-Q10), Engagement (Q11-Q15), and Learning Support (Q16-Q28). The Learning Support items include participants' perceived learning gains and decision-to-outcome understanding. Ratings range from 1 (negative) to 5 (positive), with the bar chart showing the number of respondents for each score across all questions. For consistency, the points for negative questions in the \ac{sus} scale have been reversed.}
    \label{fig:user_study-questionnaire_results}
    \vspace{-10px}
\end{figure*}

\subsection{Task-Oriented Analysis Console for Metric-driven Diagnosis}
While the graph workspace facilitates situated monitoring and interventions, learners also need metric-driven evidence to diagnose delayed effects and assess strategies. The \textit{Analysis Console} (Fig.~\ref{fig:system_overview}) provides an organized analytic space that complements visual causal tracing in the network view.

The console organizes performance indicators into three dimensions that align with common learning objectives in \ac{scm}: financial performance (Fig.~\ref{fig:system_overview}(a)), operational efficiency (Fig.~\ref{fig:system_overview}(b)), and logistics efficiency (Fig.~\ref{fig:system_overview}(c)). Each panel introduces the meaning and formula of key concepts at the top, and then presents interactive charts that visualize metric evolution over time. This design combines conceptual explanation with live evidence, helping learners interpret metrics in the context of ongoing dynamics rather than as detached summaries.
\revision{All concepts presented in the Analysis Console follow standard \ac{scm} definitions and formulas in textbooks to ensure consistency with the long-established principles. We provide the complete set of mathematical formulas and variable definitions in Appendix D.}\ques{[R2-W4]}

The console is synchronized with the active branch selected in the \textit{Timeline Tree Map}. Switching branches updates the console views accordingly, enabling learners to compare alternative trajectories using consistent metric definitions and temporal alignment. This branch-aware linkage supports diagnostic reasoning about why a strategy succeeds or fails, and helps learners relate local interventions in the graph to longer-term, system-level impacts observed in KPI trends (DG2, DG3).

\subsection{Implementation Details}
\revision{The prototype of SupplyNet is implemented using a modern web technology stack, with the frontend built in Vue 3.5 and the backend simulation core in Python environment. We developed a custom multi-agent framework where all agents are powered by OpenAI's \texttt{gpt-4o} model. Detailed information regarding the frameworks, libraries, database design, system requirements, and token cost considerations is provided in Appendix~F.}\ques{[R2-W3]} \revision{The system demo video can be found in the supplementary materials.}\ques{[R2-W6]}

\section{User Study}
We conducted a within-subjects user study to evaluate \textit{SupplyNet} against a standard baseline simulation, focusing on usability, engagement, and learning support.

\subsection{Participants}
We recruited 14 participants ($N=14$, 4 female, 10 male, aged 21--28) via university platforms. The cohort comprised 11 graduate and 3 undergraduate students majoring in business or industrial engineering. To ensure domain literacy for evaluating the learning content, all participants had foundational knowledge of \ac{scm}: 54.8\% had completed one related course, 38.5\% had completed two, and 7.7\% had completed three. Participants received \$20/h for their contribution.

\subsection{Procedure}
The study employed a within-subjects design with counterbalanced order to mitigate learning and fatigue effects. Participants completed two conditions: a baseline system (HBS Beer Game~\cite{HBSPSimulationV3_online}) and \textit{SupplyNet}.

At the beginning of each session, we introduced the study purpose and data privacy practices. In the baseline condition, participants followed the built-in instructions (approximately 15 minutes) and then played the simulation for 30 minutes. In the \textit{SupplyNet} condition, participants first received a 15-minute interactive tutorial covering the \textit{Graph Structure View}, \textit{Timeline Tree Map}, and \textit{Analysis Console}. They then completed a 30-minute task session with the goal of maximizing profit while verbalizing their thoughts using a think-aloud protocol. A mandatory 5-minute break was enforced between conditions.

After both sessions, participants completed a comparative questionnaire using five-point Likert items, followed by a semi-structured interview about their experience with \textit{SupplyNet}. Each session lasted about 90 minutes and was audio-recorded.

\subsection{Data Collection and Analysis}
We collected quantitative ratings from questionnaires and interaction logs from both conditions. For qualitative data, two authors conducted an inductive thematic analysis on the interview transcripts following Braun and Clarke’s six-step framework~\cite{braun2006using}. Codes and themes were iteratively refined through discussion; disagreements were discussed until consensus was reached to ensure that the reported themes were grounded in participant feedback.

\subsection{Quantitative Results}
The results provide evidence for \textit{SupplyNet}'s effectiveness in fulfilling design goals and outperforming the baseline.

\subsubsection*{Evaluation Against Design Goals.}
User ratings (Fig.~\ref{fig:user_study-questionnaire_results}) suggest that key design mechanisms were effective:
\begin{itemize}
     \item \textbf{Context and Feedback (DG1):} Participants rated the tiered challenge design (Q12), dynamic events (Q13), and background narratives (Q14) highly (avg. $>4.5/5$), supporting the effectiveness of contextualized gameplay. Based on interaction logs, 64\% of participants voluntarily selected the Intermediate or Advanced difficulty levels, and the average active session time was 28 minutes within the 30-minute task, suggesting sustained engagement during decision cycles.
    \item \textbf{Exploration and Comparison (DG2):} Participants reported that the \textit{Timeline Tree Map} supported exploratory analysis (Q22; 13/14 rated $\ge4$) and helped maintain engagement (Q11-Q15). Real-time profit/loss feedback (Q23) was viewed as important for guiding iterative trials. Ratings also indicate that the comparison-oriented analysis support (Q21) helped learners contrast outcomes across alternative operations.
    \item \textbf{Visual Understanding and Interpretability (DG3):} Participants rated the \textit{Analysis Console} panels (Q24-Q26) positively for clarifying system dynamics and performance breakdowns. The Reasoning Bubbles (Q27) were also rated positively by most participants, suggesting that brief agent explanations improved interpretability. Usability ratings for navigation (Q3) remained high despite the multi-view interface.
\end{itemize}

\subsubsection*{Comparative Performance.}
\begin{figure}[thp!]
    \centering
    \includegraphics[width=0.5\textwidth]{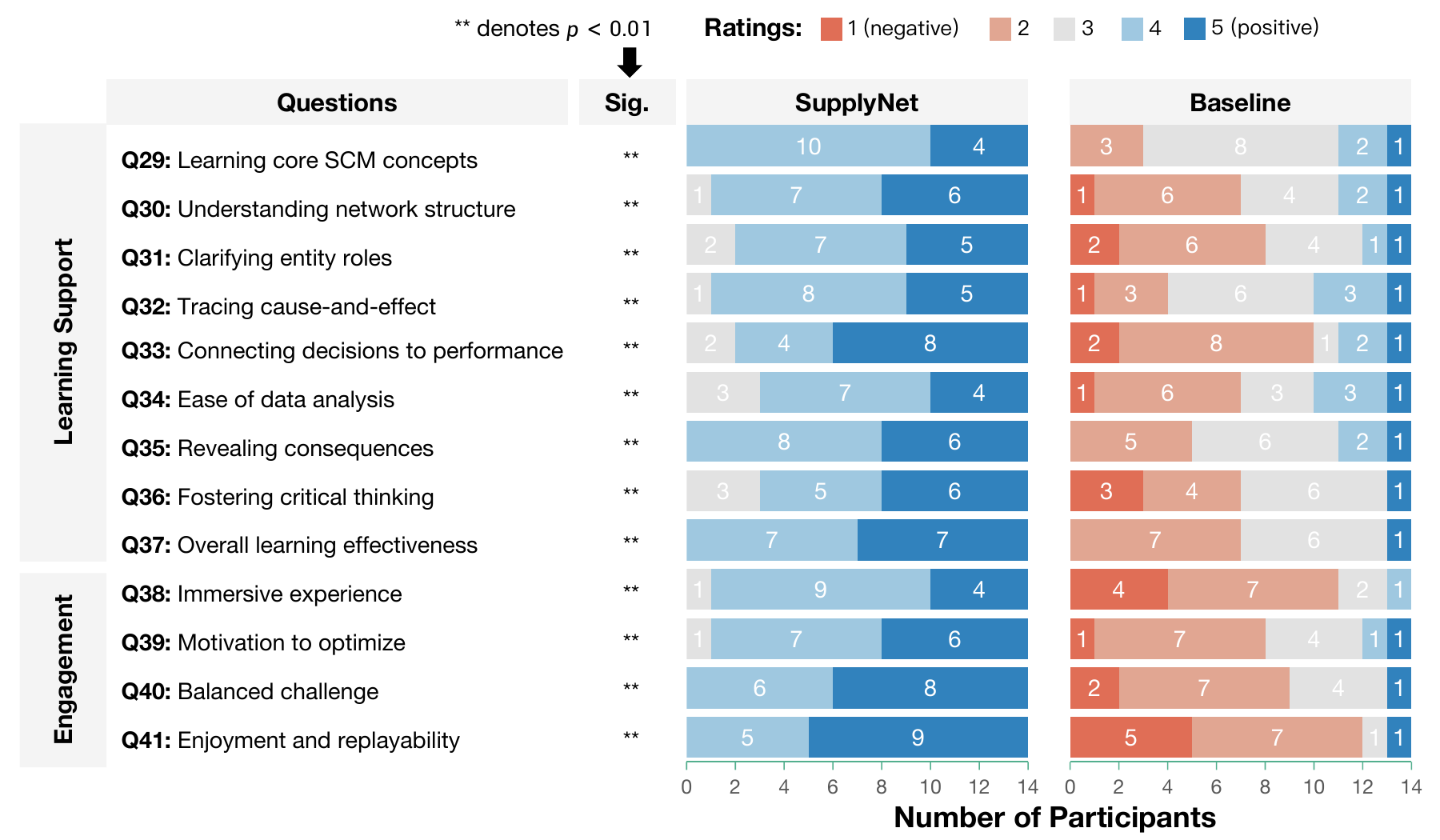}
    \vspace{-10px}
    \caption{Comparison of user ratings for \textit{SupplyNet} and the baseline system across two dimensions: Learning Support (Q29--Q37) and Engagement (Q38--Q41). The bar chart illustrates the distribution of participant ratings on a five-point scale, ranging from 1 (negative) to 5 (positive). Questions marked with a double asterisk (**) denote a statistically significant preference for \textit{SupplyNet} over the baseline (p < 0.01).}
    \label{fig:user_study-comparison_results}
    \vspace{-10px}
\end{figure}

As shown in~\ref{fig:user_study-comparison_results}, \textit{SupplyNet} showed statistically significant improvements ($p < 0.01$) over the baseline:
\begin{itemize}
    \item \textbf{Learning Support:} Participants rated \textit{SupplyNet} significantly higher in helping them learn core \ac{scm} concepts (Q29), understand network structure (Q30), and clarify entity roles in the supply chain (Q31). They also reported stronger support for reasoning about consequences, including tracing cause-and-effect relationships (Q32), connecting decisions to performance (Q33), and fostering critical thinking during play (Q36). The difference on ``connecting decisions to performance'' (Q33) was particularly notable: 13 participants rated \textit{SupplyNet} highly, compared to only 3 for the baseline. Overall perceived learning effectiveness was also higher with \textit{SupplyNet} (Q37).
    \item \textbf{Engagement:} The most pronounced difference was observed in engagement. \textit{SupplyNet} was rated as significantly more immersive (Q38), motivating (Q39), and enjoyable (Q41), whereas baseline ratings concentrated in the neutral-to-negative range.
\end{itemize}

\subsection{Qualitative Findings}
Thematic analysis revealed three recurring themes regarding the system’s value.

\subsubsection*{Theme 1: Engagement grounded in uncertainty and context.}
Participants described the event-driven simulation as bridging static theory and dynamic consequences. As P8 noted, ``It aroused my interest to see the rippling effect of sudden events.'' Several participants highly agreed that surprises and customizable disruptions enhanced replayability and inspired iterative experiments, suggesting that contextual uncertainty can sustain engagement and make feedback meaningful (DG1).

\subsubsection*{Theme 2: Counterfactual reflection enabled by branching history.}
Participants valued the ability to revisit earlier decisions and explore alternatives without irreversible penalty. They used branching for both optimization and stress testing; as P6 explained, it allowed them to ``explore extreme cases like surviving without orders.'' These accounts indicate that branching can shift attention from execution toward reflection (DG2).

\subsubsection*{Theme 3: Analytical clarity supported by coordinated visualizations.}
Participants reported that coordinated visualizations improved a sense of control by making cost drivers and trends inspectable. P2 specifically mentioned the benefit of ``self-verifying calculations'' when inspecting cost composition and formulas. Overall, the interface was perceived as turning complex network data into actionable diagnostic cues for decision making (DG3).

\section{Case Studies}
This section presents the user insights and common behavioral patterns from the students who participated in the user study. These findings illustrate how \textit{SupplyNet}’s system design features cater to diverse learning preferences and successfully foster a reciprocal relationship between learning support and engaging experience.

\subsection{Glancers vs. Diggers}\label{sec:case_study-learning_support}
While participants universally valued the learning support integrated within the system, our study revealed that the most effective forms of support differed significantly based on the user's level of expertise. This suggests that a single approach to information design is insufficient for a diverse user base.

For participants at the beginner level (P4, P8-10, P14), support was most effective when delivered through immediate, intuitive visual cues. They preferred to make at-a-glance assessments rather than delve into data. For example, P8 explained how he relied on simple color indicators, noting, ``\textit{Those visual encodings help me grasp a quick, intuitive understanding of the environmental changes at each time point.}'' Similarly, P4 frequently referred to the \textit{Timeline Tree Map}, where color saturation illustrated performance over time. He found this feature highly effective, stating, ``\textit{It helps evaluate my business performance without needing to dive into complex data and naturally motivates me to reflect on how my past history shaped my current status.}'' For these users, the \textit{Analysis Console} was consulted only when the visual information was insufficient or confusing. As P8 admitted, ``\textit{Despite the many useful information in the analysis console, it makes me feel overwhelmed sometimes. I am willing to investigate it when the visuals cannot be understood intuitively.}''

In contrast, participants at an advanced level (P3, P12) gravitated towards the \textit{Analysis Console}, finding enjoyment and a sense of mastery in engaging directly with the data. For this group, the visuals were secondary to the precise figures and formulas. P12, for instance, valued the ability to perform calculations, dissect trends, and understand the exact definitions of key metrics. He explained that the detailed analysis was the primary source of his engagement: ``\textit{This deep analysis provided me a sense of empowerment and mastery, knowing every detail about how profits come and the performance of me and every one of my suppliers.}''

This clear distinction in preference underscores the need for a multi-layered support system that not only provides immediate visual intuition to guide beginner students but also offers the analytical support to challenge advanced learners.

\subsection{Which Comes First: The Fun or the Facts?}\label{sec:case_study-reciprocative_relationship}
We observed two primary pathways into a virtuous cycle of learning and enjoyment. Some participants were first drawn in by engaging visuals which served as a gateway to deeper analysis, while others found enjoyment directly in the act of learning by mastering the analytical tools and data.

For some participants (P5-6, P8-10, P14), visually engaging features served as a gateway to the more complex analytical tools. For instance, P5 was initially drawn to the vibrant colors in the \textit{Analysis Console}. This visual curiosity motivated him to examine the underlying figures and understand the insights behind them. Through a few rounds of play, he became familiar with complex cost compositions and professional metrics. Similarly, P6 was first attracted to the \textit{Timeline Tree Map}, playfully creating multiple timelines to observe their consequences. The resulting color variations on the timelines motivated him to think more critically. In these cases, features designed for exploration and immediate feedback served as a bridge, naturally guiding users toward analytical learning.

For other participants(P2-3, P12), the act of learning itself becomes the primary driver of engagement. This was particularly evident among those with science or engineering backgrounds, who were immediately drawn to the \textit{Analysis Console} and the wealth of information it provided. Engagement was achieved through mastery and understanding. P12 was initially unsure of his grasp of ``inventory turns''. He then used the provided formula to manually calculate the value from the data shown on the interface. After confirming his self-calculated number matched the one on the chart, he reported a profound sense of joy. As he articulated, \textit{``Understanding in detail gives me a sense of achievement, and it further leads to joy.''}

\section{Discussion}
In this section, we discuss the key findings from our case study and their implications for designing effective educational simulations. First, we explore the divergence in user preferences for learning support, attributing this to the interplay between prior domain knowledge and cognitive load. Then we will examine the reciprocal relationship our system successfully creates between engagement and learning, and discuss how this dual-pathway approach can guide the future development of adaptive learning systems. We will conclude by discussing the limitations of our system and outlining future research directions.

\subsection{Different preferences for learning support}
The observed divergence in preferences for intuitive versus analytical learning support in~\cref{sec:case_study-learning_support} can be attributed to two interconnected factors: the participants' background knowledge and their cognitive load.

The level of prior domain knowledge determines the learning behaviors. Participants who had taken more courses possessed a more systematic understanding of supply chain. Thus, they are capable of interpreting inter-related variables and engaging with the \textit{Analysis Console} directly without feeling disoriented. However, participants at a beginner level lacked the knowledge foundation, making them more likely to feel overwhelmed by the density of graphs and metrics presented in the console.

This difference in background knowledge directly correlates with the second factor: cognitive load. For advanced users, their expertise automated the interpretation of basic concepts, freeing up cognitive resources to focus on higher-level strategic thinking and data analysis. This made the analytical process feel empowering rather than burdensome. In contrast, novices had to process every piece of information from terminology to graphical representations, which rapidly consumed their cognitive capacity and led to feelings of being overwhelmed.

This distinction highlights the need for progressive disclosure of information and analytical learning support. To avoid overwhelming beginner-level users, the system should initially offer a streamlined, visual interface that minimizes information density and fosters motivation. As users gain expertise, they can be introduced to the richer analytical tools and data granularity that advanced learners require to achieve mastery and make the simulation a more robust training environment. This adaptive approach helps create a scaffold that guides all users from initial engagement toward deeper analytical understanding.

\subsection{Reciprocal Relationships between Engagement and Learning Support}

Engagement and learning support are often treated as distinct, sometimes competing, dimensions in educational design. Some systems, such as Capitalism Lab prioritize engagement through rich visual stimuli and playful design, but at the expense of substantive learning. Others, like the baseline system in this study, focus heavily on learning by presenting dense data and complex analytical tasks, which can sacrifice a fun and engaging user experience. 

In contrast, the findings in~\cref{sec:case_study-reciprocative_relationship} demonstrates that our system builds a successful reciprocal relationship between these two elements, where each one strengthens the other. We observed two primary pathways into this positive feedback loop. Some participants were initially captivated by the visual appeal and interactive elements, which served as an effective gateway to exploring its deeper educational content. For them, engagement led to learning. Others found engagement through the act of learning itself, deriving a sense of mastery and intellectual satisfaction from analyzing data and mastering the simulation’s mechanics. For them, learning was inherently engaging.

This dual pathway highlights that an effective educational simulation should provide entry points for both types of users. It must cater to those drawn in by playful interaction and visual appeal, as well as those motivated by the intellectual satisfaction of analysis and mastery. By doing so, all students, regardless of their starting point, can enter a rewarding cycle where engagement fuels learning, and successful learning, in turn, deepens engagement.

\begin{revisionnblock}
\subsection{Generalizability Across Subjects}\label{sec:genralizavbility}

Although SupplyNet is designed and evaluated in the context of supply chain management, its broader contribution is not limited to the direct reuse of SCM-specific entities, actions, or metrics. Rather, the system embodies a reusable interaction pattern for subjects that can be modeled as dynamic relational systems. Such subjects typically involve multiple connected entities, local decisions or disturbances, delayed propagation of effects, and trade-offs that learners need to diagnose over time. Under this abstraction, SupplyNet separates a reusable layer from a domain-specific layer. The reusable layer consists of graph-based state representation, contextualized subgraph retrieval and role/task framing, event injection, temporal snapshots, branchable histories, coordinated visual views, and metric-driven reflection. The domain-specific layer consists of node and edge semantics, valid learner actions, agent heuristics, event types, transition rules, and performance indicators. Therefore, the framework is most transferable to subjects that share three learning properties: knowledge emerges from interactions among multiple entities, learners benefit from testing interventions over time, and learning objectives can be expressed through interpretable state changes and performance indicators.

This separation clarifies how each view and interaction technique can be adapted beyond SCM. The onboarding guide functions as a role-and-complexity calibration mechanism: company roles, challenge levels, and narrative missions can be replaced by domain roles such as public-health officers, urban planners, ecosystem managers, or grid operators, while difficulty can be controlled through network size, uncertainty level, number of controllable policies, or severity of external events. The Graph Structure View functions as the main inspection-and-intervention workspace. Its nodes and edges can be redefined from companies and logistics links to regions and mobility flows, intersections and roads, species and ecological interactions, or generators and transmission lines. Node and edge inspection remains a details-on-demand interaction, but the inspected attributes change from inventory, capacity, backlog, and lead time to domain-specific states such as infection prevalence, congestion level, population size, habitat quality, energy load, or system reliability.

The Node Editing Panel and event palette can be preserved as tools for controlled perturbation. Instead of adding suppliers or injecting storms and shortages, learners may add transportation links, allocate medical resources, restore habitats, adjust grid capacity, or introduce policy shocks and environmental disturbances. The Reasoning Bubble can also be generalized as an interpretability layer. In human-centered domains, it can explain the decision logic of simulated actors such as hospitals, commuters, local authorities, or households. In non-human domains, such as ecology or energy systems, it can instead summarize the local rule or model mechanism that explains why a state changed. The Timeline Tree Map provides a branchable history for counterfactual reasoning, allowing learners to return to a prior state, alter an intervention, and compare alternative futures. The Analysis Console remains a metric-aligned reflection space, where SCM dashboards for financial, operational, and logistics performance are replaced with indicators aligned with the target subject and learning objectives.

Building on this component-level mapping, Table~\ref{tab:cross_subject_instantiations} illustrates how the abstract framework can be instantiated in representative subjects by replacing the graph schema, learner interventions, contextual events, and metric dashboards while preserving the same exploratory workflow.
\end{revisionnblock}

\begin{table*}[t]
\begingroup
\centering
\small
\setlength{\tabcolsep}{4pt}
\renewcommand{\arraystretch}{1.15}
\captionsetup{labelfont={color=black},textfont={color=black}}
\caption{Representative instantiations of the generalized SupplyNet framework beyond SCM.}
\label{tab:cross_subject_instantiations}
\begin{tabular}{p{0.14\linewidth} p{0.24\linewidth} p{0.27\linewidth} p{0.27\linewidth}}
\hline
\textbf{Subject} & \textbf{Graph schema} & \textbf{Learner interventions and events} & \textbf{Metrics and learning tasks} \\
\hline
Public health and epidemiology &
Nodes represent regions, hospitals, or population groups; edges represent mobility, contact, or patient-transfer links. &
Learners allocate vaccines, testing resources, hospital beds, or mobility restrictions. Events may include new variants, public gatherings, seasonal changes, or vaccine shortages. &
Dashboards can summarize infection prevalence, hospital utilization, intervention cost, mortality, and regional equity. Learning tasks include understanding disease propagation, delayed intervention effects, resource-allocation trade-offs, and policy timing. \\

Urban planning and transportation &
Nodes represent districts, intersections, stations, or public facilities; edges represent roads, transit lines, commuting flows, or accessibility links. &
Learners adjust signal timing, transit frequency, zoning policies, road capacity, or emergency routing. Events may include accidents, weather disruptions, population growth, or infrastructure failures. &
Dashboards can summarize travel time, congestion, accessibility, emissions, budget, and service reliability. Learning tasks include identifying network bottlenecks, reasoning about induced demand, and comparing efficiency-equity trade-offs. \\

Ecology and environmental management &
Nodes represent species, habitats, water bodies, or resource pools; edges represent predation, competition, migration, or nutrient flows. &
Learners introduce conservation policies, harvest limits, habitat restoration, or pollution controls. Events may include droughts, invasive species, wildfires, or climate shocks. &
Dashboards can summarize population stability, biodiversity, resource availability, extinction risk, recovery time, and intervention cost. Learning tasks include exploring trophic cascades, resilience, delayed ecological feedback, and conservation-resource trade-offs. \\

Energy systems and smart grids &
Nodes represent generators, substations, storage units, consumers, or regions; edges represent transmission lines and energy flows. &
Learners dispatch generators, schedule storage, adjust demand-response policies, or repair failures. Events may include heat waves, renewable fluctuations, transmission failures, or price shocks. &
Dashboards can summarize reliability, load shedding, generation cost, carbon emissions, renewable utilization, and peak demand. Learning tasks include understanding supply-demand balancing, cascading failures, renewable intermittency, and cost-emission trade-offs. \\
\hline
\end{tabular}
\endgroup
\end{table*}

\begin{revisionnblock}
The instantiations in Table~\ref{tab:cross_subject_instantiations} suggest a practical adaptation workflow. First, designers identify whether the target subject can be represented as a dynamic relational system and define its graph schema, including entities, relationships, state variables, and controllable actions. Second, they encode domain rules or agent heuristics that constrain how the system evolves and how autonomous actors respond. Third, they define contextual events that perturb local states and make propagation mechanisms observable. Fourth, they replace SCM-specific dashboards with metrics aligned with the target learning objectives. Finally, they preserve the same exploratory workflow: role-based onboarding, graph-based inspection and intervention, event-driven feedback, branchable counterfactual comparison, and dashboard-based reflection.

This abstraction also clarifies the boundary of generalization. SupplyNet is less suitable for learning tasks dominated by static memorization or one-shot procedural training unless those tasks can be embedded in a meaningful relational process. Cross-subject deployment would still require collaboration with domain instructors to validate the graph schema, transition rules, agent behavior, event design, and assessment metrics. The generalizable contribution of SupplyNet is thus not a universal plug-and-play simulator, but a reusable design framework for transforming domain-specific dynamic systems into interactive, visual, and counterfactual learning environments.
\end{revisionnblock}\quess{Q1}

\subsection{Limitations and Future Work}

The proposed framework has certain limitations that lead to future research directions.

\textbf{Further eliciting the potential of \acp{llm} in educational simulations.}
Beyond their role as business managers, the capabilities of \acp{llm} can be harnessed in other aspects of the simulation to enhance engagement and learning support. One significant potential lies in simulating complex, human-like interactions such as negotiations, leveraging capacity for understanding and reasoning. Furthermore, \acp{llm} can be integrated with the \textit{Analysis Console} to act as an on-demand analytical assistant. This contextual just-in-time aid would bridge the gap between observing data and understanding core \ac{scm} concepts, making the learning process more intuitive and effective. By leveraging these capabilities, future educational simulations can become more interactive, responsive, and interesting. \revision{Beyond supply chain education, the framework's interaction design---gamified onboarding, branching timeline exploration, and task-oriented analysis dashboards---could potentially be adapted to other subjects involving dynamic systems with interdependent decision-making, such as public health policy or urban planning, provided that domain-specific graph schemas, agent heuristics, and performance metrics are defined.}\ques{[R2-W2]}

\textbf{Benchmarking user performance in educational simulations.}
The simulation is designed to motivate students to explore alternative strategies and explore ``what-if'' scenarios via \textit{Timeline Tree Map}. However, the system currently offers limited support in helping students benchmark their decisions. For instance, students may maximize profits through ``trial and errors'', but lack insights into optimal outcomes they could achieve. To enhance learning support, the system could integrate optimization algorithms to calculate the upper and lower bounds of potential business performance, motivating students with a clearer understanding of achievable results.

\textbf{Reducing computational cost.} The computational complexity of the framework increases significantly with the number of agents. Specifically, the complexity scales as $T(M\cdot N)$, where $M$ represents the number of stages and $N$ represents the total number of agents in the framework. The computational cost for simulation at the current scale is acceptable. However, this could become a concern in the custom mode, where users have the freedom to construct networks at a much larger scale. It may lead to prohibitive computational costs and long waiting times, making the simulation unaffordable. To mitigate this, techniques such as quantization, model compression, or parallel computing may be required to reduce the computational costs.

\section{Conclusion}

This work focused on the challenge of creating educational simulations for \ac{scm} that are often disengaging and fail to provide sufficient learning support. Based on insights from our formative study, we developed \textit{SupplyNet}, a
gamified visual simulation system built on a contextual graph-based LLM multi-agent framework that models complex, interdependent supply chain dynamics. The \textit{SupplyNet} interface visualizes and transforms the simulated supply chain into an interactive
learning environment, where students can experiment with alternative strategies, and conduct analysis on business performance metrics. Our evaluation demonstrated that the proposed system significantly enhanced the learning support for students while promoting a fun and engaging learning experience. Furthermore, it successfully creates a reciprocal relationship between engagement and learning, effectively supporting students at different learning stages.

\section*{Acknowledgement}
This work is supported by RGC GRF Grant 16210321.



\bibliographystyle{elsarticle-harv}
\bibliography{Reference/Ref-Computer_Science,Reference/Ref-Education,Reference/Ref-Supply_Chain_Management}

@String{Computing = "Computing" }

@String{Computer = "{IEEE} Computer" }

@String{Springer = "Springer-Verlag" }

@inproceedings{schade2023mapuncover,
  title={{MapUncover}: Fostering spatial exploration through gamification in mobile map apps},
  author={Schade, Eve and Savino, Gian-Luca and Niess, Jasmin and Sch{\"o}ning, Johannes},
  booktitle={Proceedings of the 2023 CHI Conference on Human Factors in Computing Systems},
  year={2023},
  doi = {10.1145/3544548.3581428},
}

@inproceedings{du2024careersim,
  title={{CareerSim}: Gamification design leveraging {LLMs} for career development reflection},
  author={Du, Wantong and Zhu, Zhiying and Xu, Xinhui and Che, Haoyuan and Chen, Shi},
  booktitle={Extended abstracts of the CHI conference on Human Factors in Computing Systems},
  year={2024},
  doi = {10.1145/3613905.3650928},
}

@inproceedings{mohaddesi2022trust,
  title={To trust or to stockpile: Modeling human-simulation interaction in supply chain shortages},
  author={Mohaddesi, Omid and Griffin, Jacqueline and Ergun, Ozlem and Kaeli, David and Marsella, Stacy and Harteveld, Casper},
  booktitle={Proceedings of the 2022 CHI Conference on Human Factors in Computing Systems},
  year={2022},
  doi = {10.1145/3491102.3502089},
}

@inproceedings{kirchner2024outplay,
  title={Outplay your weaker self: A mixed-methods study on gamification to overcome procrastination in academia},
  author={Kirchner-Krath, Jeanine and Schmidt-Kraepelin, Manuel and Sch{\"o}bel, Sofia and Ullrich, Mathias and Sunyaev, Ali and Von Korflesch, Harald FO},
  booktitle={Proceedings of the 2024 CHI Conference on Human Factors in Computing Systems},
  year={2024},
  doi = {10.1145/3613904.3642048},
}

@inproceedings{feger2019gamification,
author = {Feger, Sebastian S. and Dallmeier-Tiessen, S\"{u}nje and Wo\'{z}niak, Pawe\l{} W. and Schmidt, Albrecht},
title = {Gamification in Science: A Study of Requirements in the Context of Reproducible Research},
year = {2019},
isbn = {9781450359702},
doi = {10.1145/3290605.3300690},
booktitle = {Proceedings of the 2019 CHI Conference on Human Factors in Computing Systems},
pages = {1–14},
numpages = {14},
location = {Glasgow, Scotland Uk},
series = {CHI '19}
}

@inproceedings{zhang2024see,
  title={See widely, think wisely: Toward designing a generative multi-agent system to burst filter bubbles},
  author={Zhang, Yu and Sun, Jingwei and Feng, Li and Yao, Cen and Fan, Mingming and Zhang, Liuxin and Wang, Qianying and Geng, Xin and Rui, Yong},
  booktitle={Proceedings of the 2024 CHI Conference on Human Factors in Computing Systems},
  year={2024},
  doi = {10.1145/3613904.3642545},
}

@inproceedings{zhang2023visar,
  title={{VISAR}: A human-ai argumentative writing assistant with visual programming and rapid draft prototyping},
  author={Zhang, Zheng and Gao, Jie and Dhaliwal, Ranjodh Singh and Li, Toby Jia-Jun},
  booktitle={Proceedings of the 36th Annual ACM Symposium on User Interface Software and Technology},
  year={2023},
  doi = {10.1145/3586183.3606800},
}

@inproceedings{li2024econagent,
  title={{EconAgent}: Large Language Model-Empowered Agents for Simulating Macroeconomic Activities},
  author={Li, Nian and Gao, Chen and Li, Mingyu and Li, Yong and Liao, Qingmin},
  booktitle={Proceedings of the 62nd Annual Meeting of the Association for Computational Linguistics (Volume 1: Long Papers)},
  year={2024},
  doi = {10.18653/v1/2024.acl-long.829},
}

@inproceedings{park2022social,
  title={Social simulacra: Creating populated prototypes for social computing systems},
  author={Park, Joon Sung and Popowski, Lindsay and Cai, Carrie and Morris, Meredith Ringel and Liang, Percy and Bernstein, Michael S},
  booktitle={Proceedings of the 35th Annual ACM Symposium on User Interface Software and Technology},
  year={2022},
  doi = {10.1145/3526113.3545616},
}

@article{gao2023s3,
  title={S3: Social-network simulation system with large language model-empowered agents},
  author={Gao, Chen and Lan, Xiaochong and Lu, Zhihong and Mao, Jinzhu and Piao, Jinghua and Wang, Huandong and Jin, Depeng and Li, Yong},
  journal={arXiv preprint arXiv:2307.14984},
  year={2023},
  doi = {10.2139/ssrn.4607026},
}

@inproceedings{zhao2024competeai,
author = {Zhao, Qinlin and Wang, Jindong and Zhang, Yixuan and Jin, Yiqiao and Zhu, Kaijie and Chen, Hao and Xie, Xing},
title = {CompeteAI: understanding the competition dynamics of large language model-based agents},
year = {2024},
publisher = {JMLR.org},
booktitle = {Proceedings of the 41st International Conference on Machine Learning},
articleno = {2526},
numpages = {16},
location = {Vienna, Austria},
series = {ICML'24},
doi = {10.5555/3692070.3694596}
}

@inproceedings{lin2024strategic,
  title={Strategic Collusion of {LLM} Agents: Market Division in Multi-Commodity Competitions},
  author={Lin, Ryan Y and Ojha, Siddhartha and Cai, Kevin and Chen, Maxwell},
  booktitle={Language Gamification-NeurIPS 2024 Workshop},
  year={2024}
}

@misc{han2023guinea,
      title={"Guinea Pig Trials" Utilizing {GPT}: A Novel Smart Agent-Based Modeling Approach for Studying Firm Competition and Collusion}, 
      author={Xu Han and Zengqing Wu and Chuan Xiao},
      year={2024},
      eprint={2308.10974},
      archivePrefix={arXiv},
      primaryClass={cs.AI}
}

@inproceedings{yuzhe2025twinmarket,
  title={{TwinMarket}: A Scalable Behavioral and Social Simulation for Financial Markets},
  author={Yang, Yuzhe and Zhang, Yifei and Wu, Minghao and Zhang, Kaidi and Zhang, Yunmiao and Yu, Honghai and Hu, Yan and Wang, Benyou},
  booktitle={ICLR 2025 Workshop on World Models: Understanding, Modelling and Scaling},
  year={2025}
}

@article{quan2024invagent,
  title={{InvAgent}: A Large Language Model based {Multi-Agent} System for Inventory Management in Supply Chains},
  author={Quan, Yinzhu and Liu, Zefang},
  journal={CoRR},
  year={2024}
}

@inproceedings{jiang2024personallm,
  title={{PersonaLLM}: Investigating the Ability of Large Language Models to Express Personality Traits},
  author={Jiang, Hang and Zhang, Xiajie and Cao, Xubo and Breazeal, Cynthia and Roy, Deb and Kabbara, Jad},
  booktitle={Findings of the Association for Computational Linguistics: NAACL 2024},
  year={2024},
  doi = {10.18653/v1/2024.findings-naacl.229},
}

@inproceedings{choi2024picle,
  title={{PICLe}: eliciting diverse behaviors from large language models with persona in-context learning},
  author={Choi, Hyeong Kyu and Li, Yixuan},
  booktitle={Proceedings of the 41st International Conference on Machine Learning},
  year={2024}
}

@inproceedings{2023gvqa,
    author = {Song, Sicheng and Chen, Juntong and Li, Chenhui and Wang, Changbo},
    title = {{GVQA}: Learning to Answer Questions about Graphs with Visualizations via Knowledge Base},
    year = {2023},
    booktitle = {Proceedings of the 2023 CHI Conference on Human Factors in Computing Systems},
  doi = {10.1145/3544548.3581067},
}

@incollection{shneiderman2003eyes,
  title={The eyes have it: A task by data type taxonomy for information visualizations},
  author={Shneiderman, Ben},
  booktitle={The craft of information visualization},
  pages={364--371},
  year={2003},
  publisher={Elsevier},
  doi = {10.1016/b978-155860915-0/50046-9},
}

@ARTICLE{song2024gvvst,
  author={Song, Sicheng and Zhang, Yipeng and Lin, Yanna and Qu, Huamin and Wang, Changbo and Li, Chenhui},
  journal=TVCG, 
  title={GVVST: Image-Driven Style Extraction From Graph Visualizations for Visual Style Transfer}, 
  year={2025},
  volume={31},
  number={9},
  pages={5975-5989},
  doi={10.1109/TVCG.2024.3485701}}

@ARTICLE{song2026vizdefender,
  author={Song, Sicheng and Zhang, Yanjie and Chen, Zixin and Qu, Huamin and Wang, Changbo and Li, Chenhui},
  journal={IEEE Transactions on Visualization and Computer Graphics}, 
  title={VizDefender: Unmasking Visualization Tampering Through Proactive Localization and Intent Inference}, 
  year={2026},
  volume={32},
  number={6},
  pages={4720-4730},
  doi={10.1109/TVCG.2026.3694448}}

@article{zhang2026adapt,
  title={AdaPT: Adaptive Lesson Plan Transformer for Cross-Regional and Differentiated Instruction},
  author={Zhang, Yanjie and Zhu, Jiajun and Wu, Minyu and Qu, Huamin and Song, Sicheng},
  journal={arXiv preprint arXiv:2606.17633},
  year={2026}
}

@ARTICLE{song2023vividgraph,
  author={Song, Sicheng and Li, Chenhui and Sun, Yujing and Wang, Changbo},
  journal={IEEE Transactions on Visualization and Computer Graphics}, 
  title={VividGraph: Learning to Extract and Redesign Network Graphs From Visualization Images}, 
  year={2023},
  volume={29},
  number={7},
  pages={3169-3181},
  doi={10.1109/TVCG.2022.3153514}}

@ARTICLE{song2024graphdecoder,
  author={Song, Sicheng and Li, Chenhui and Li, Dong and Chen, Juntong and Wang, Changbo},
  journal={IEEE Transactions on Visualization and Computer Graphics}, 
  title={GraphDecoder: Recovering Diverse Network Graphs From Visualization Images via Attention-Aware Learning}, 
  year={2024},
  volume={30},
  number={7},
  pages={3074-3088},
  doi={10.1109/TVCG.2022.3225554}}

@inproceedings{chen2025unmasking,
  title={Unmasking deceptive visuals: Benchmarking multimodal large language models on misleading chart question answering},
  author={Chen, Zixin and Song, Sicheng and Shum, Kashun and Lin, Yanna and Sheng, Rui and Wang, Weiqi and Qu, Huamin},
  booktitle={Proceedings of the 2025 Conference on Empirical Methods in Natural Language Processing},
  pages={13767--13800},
  year={2025}
}

@article{pfeffer2002end,
  title={The end of business schools? Less success than meets the eye},
  author={Pfeffer, Jeffrey and Fong, Christina T},
  journal={Academy of management learning \& education},
  volume={1},
  number={1},
  pages={78--95},
  year={2002},
  publisher={Academy of Management Briarcliff Manor, NY 10510},
  doi = {10.5465/amle.2002.7373679},
}

@article{pettigrew2016guest,
  title={From the guest editors: The legitimacy and impact of business schools—Key issues and a research agenda},
  author={Pettigrew, Andrew and Starkey, Ken},
  journal={Academy of Management Learning \& Education},
  volume={15},
  number={4},
  pages={649--664},
  year={2016},
  publisher={Academy of Management Briarcliff Manor, NY},
  doi = {10.5465/amle.2016.0296},
}

@article{dominguez2013gamifying,
  title={Gamifying learning experiences: Practical implications and outcomes},
  author={Dom{\'\i}nguez, Adri{\'a}n and Saenz-de-Navarrete, Joseba and De-Marcos, Luis and Fern{\'a}ndez-Sanz, Luis and Pag{\'e}s, Carmen and Mart{\'\i}nez-Herr{\'a}iz, Jos{\'e}-Javier},
  journal={Computers \& education},
  volume={63},
  pages={380--392},
  year={2013},
  publisher={Elsevier},
  doi = {10.1016/j.compedu.2012.12.020},
}

@article{braun2006using,
  title={Using thematic analysis in psychology},
  author={Braun, Virginia and Clarke, Victoria},
  journal={Qualitative research in psychology},
  volume={3},
  number={2},
  pages={77--101},
  year={2006},
  publisher={Taylor \& Francis},
  doi = {10.1191/1478088706qp063oa},
}

@article{roese1997counterfactual,
  title={Counterfactual thinking.},
  author={Roese, Neal J},
  journal={Psychological bulletin},
  volume={121},
  number={1},
  pages={133},
  year={1997},
  publisher={American Psychological Association},
  doi = {10.1037/0033-2909.121.1.133},
}

@article{scaife1996external,
  title={External cognition: how do graphical representations work?},
  author={Scaife, Mike and Rogers, Yvonne},
  journal={International journal of human-computer studies},
  volume={45},
  number={2},
  pages={185--213},
  year={1996},
  publisher={Elsevier},
  doi = {10.1006/ijhc.1996.0048},
}

@ARTICLE{chen2026mis,
  author={Chen, Zixin and Zeng, Yuhang and Song, Sicheng and Lin, Yanna and Xu, Xian and Qu, Huamin and Xia, Meng},
  journal={IEEE Transactions on Visualization and Computer Graphics}, 
  title={VizQStudio: Iterative Visualization Literacy MCQs Design with Simulated Students}, 
  year={2026},
  volume={},
  number={},
  pages={1-18},
  doi={10.1109/TVCG.2026.3695959}}

@string{POM = "Production and Operations Management"}

@string{IJPE = "International Journal of Production Economics"}

@string{IJPR = "International Journal of Production Research"}

@string{OR = "Operation Research"}

@string{DS = "Decision Sciences"}

@misc{skilldynamics2023beer,
    title = {Supply Chain Beer Game - Virtual Simulation},
    author = {{Skill Dynamics}},
    year = {2023},
    url = {https://skilldynamics.com/supply-chain-beer-game/},
    organization = {Skill Dynamics}
}

@article{stadtler2014supply,
  title={Supply chain management: An overview},
  author={Stadtler, Hartmut},
  journal={Supply Chain Management and Advanced Planning: Concepts, Models, Software, and Case Studies},
  pages={3--28},
  year={2014},
  publisher={Springer},
  doi = {10.1007/978-3-540-74512-9_2},
}

@article{xue2005agent,
  title={An agent-based framework for supply chain coordination in construction},
  author={Xue, Xiaolong and Li, Xiaodong and Shen, Qiping and Wang, Yaowu},
  journal={Automation in Construction},
  volume={14},
  number={3},
  pages={413--430},
  year={2005},
  publisher={Elsevier},
  doi = {10.1016/j.autcon.2004.08.010},
}

@article{swaminathan1998modeling,
  title={Modeling supply chain dynamics: A multiagent approach},
  author={Swaminathan, Jayashankar M and Smith, Stephen F and Sadeh, Norman M},
  journal=DS,
  volume={29},
  number={3},
  pages={607--632},
  year={1998},
  publisher={Wiley Online Library},
  doi = {10.1111/j.1540-5915.1998.tb01356.x},
}

@article{kotecha2024leveraging,
  title={Leveraging Graph Neural Networks and Multi-Agent Reinforcement Learning for Inventory Control in Supply Chains},
  author={Kotecha, Niki and Chanona, Antonio del Rio},
  journal={arXiv preprint arXiv:2410.18631},
  year={2024},
  doi = {10.1016/j.compchemeng.2025.109111},
}

@article{liu2022multi,
  title={Multi-agent deep reinforcement learning for multi-echelon inventory management},
  author={Liu, Xiaotian and Hu, Ming and Peng, Yijie and Yang, Yaodong},
  journal=POM,
  pages={10591478241305863},
  year={2022},
  publisher={SAGE Publications Sage CA: Los Angeles, CA},
  doi = {10.2139/ssrn.4262186},
}

@article{persson2002performance,
  title={Performance simulation of supply chain designs},
  author={Persson, Fredrik and Olhager, Jan},
  journal=IJPE,
  volume={77},
  number={3},
  pages={231--245},
  year={2002},
  publisher={Elsevier},
  doi = {10.1016/s0925-5273(00)00088-8},
}

@article{datta2011information,
  title={Information sharing and coordination mechanisms for managing uncertainty in supply chains: a simulation study},
  author={Datta, Partha Priya and Christopher, Martin G},
  journal=IJPR,
  volume={49},
  number={3},
  pages={765--803},
  year={2011},
  publisher={Taylor \& Francis},
  doi = {10.1080/00207540903460216},
}

@article{cannas2024artificial,
  title={Artificial intelligence in supply chain and operations management: a multiple case study research},
  author={Cannas, Violetta Giada and Ciano, Maria Pia and Saltalamacchia, Mattia and Secchi, Raffaele},
  journal=IJPR,
  volume={62},
  number={9},
  pages={3333--3360},
  year={2024},
  publisher={Taylor \& Francis},
  doi = {10.1080/00207543.2023.2232050},
}

@misc{HBSPSimulationV3_online,
  author    = {{Harvard Business School}},
  title     = {Global Supply Chain Management Simulation V3},
  year      = {2023},
  publisher = {Harvard Business Publishing Education},
  url       = {https://hbsp.harvard.edu/product/7908-HTM-ENG?Ntt=supply+chain+simulation},
  urldate   = {2025-07-10}
}

\clearpage
\newpage
\onecolumn
\appendix

\label{app1}

\section{Details of Semi-Structured Interview for the Formative Study} \label{appendix:Appendix A}  

\subsection{Interview Questions to Instructors}
\textbf{PART I: Background and Experience}
\textbf{Teaching Background}
\begin{itemize}
    \item Which specific supply chain courses have you taught?
    \item Based on your experience, which concepts or knowledge points in supply chain management are the most difficult for students to understand and master?
    \item What do you think are the main reasons for these difficulties? Is it because the concepts are too abstract, or due to a lack of practical environment?
    \item What methods are you currently using (\eg, case studies, classroom discussions, guest lectures) to help students address the limitations of textbook knowledge? Do you think practicing in a simulation system can help overcome these difficulties?
    \item In your attitudes toward university education for business students, do you believe students need more hands-on training in a realistic environment, either for the purpose of understanding concepts or for applying them in practice?
\end{itemize}
\begin{itemize}
    \item Which simulation tools have you used in your teaching?
    \item (If the answer to the previous question is ``Yes'')
    \item What is this simulation system? What functions does it support?
    \item What was the learning objective to achieve by using the educational supply chain simulation system?
    \item How did you integrate the simulation system into your courses (\eg, in-class practice/tutorial/after-class exercise)
    \item What determines whether you choose to use a simulation for a particular concept?
    \item Did it efficiently help you achieve the learning objective?
\end{itemize}

\textbf{PART II: Evaluation of the Existing Beer Game System}
\begin{itemize}
    \item Based on your impression on this system, What value or advantages do you think such tools bring to teaching? Which concepts can they help students intuitively understand? (\eg, safety stock, reorder points, lead time, etc.)
    \item if we take a longer-term perspective, what do you think are the obvious limitations of this system? Or, in other words, which important supply chain knowledge points does it fail to cover?
    \item For a simple exercise aimed solely at helping students practice the fundamentals of `inventory management,' do you think it is sufficient? Why?
    \item In your opinion, how much gap exists between this `single-chain' model and a real-world supply chain? Does this gap affect the development of students' skills or fail to meet your teaching needs?
\end{itemize}

\textbf{PART III: Graph-Based Modeling of Supply Chain Networks}
\textbf{From Limitations to Complexity Needs}
\begin{itemize}
    \item You just mentioned the limitations of the single-chain model. In the real world, supply chains are rarely linear; instead, they form complex networks with multiple suppliers, production facilities, and distribution channels. How necessary do you think it is for students to engage in simulations within a networked environment for their learning?
\end{itemize}
\begin{itemize}
    \item Imagine a more powerful simulator: students can make decisions on a 'graph' (network) with multiple companies. Each has their own supply and demand connections. The students can even design or modify the network themselves. In your opinion, what kinds of concepts that are hard to explain through textbooks could be better understood with such a model?
    \item For instance, concepts like `the propagation of the bullwhip effect across a network,' `risk diffusion caused by the disruption of a node (\eg, factory shutdowns or port closures),' or `how to balance inventory across multiple warehouses'—do? Do you think these could be better addressed in such a model?
\end{itemize}

\begin{itemize}
    \item In a network-based or graph-based simulator, what types of decisions would you most want students to experience and practice?
    \item (Providing options for inspiration)``Would it be decisions about network structure (\eg, where to build warehouses, which suppliers to choose), or operational decisions (\eg, designing multi-echelon inventory strategies, planning transportation routes)?''
    \item In your view, what key features should an ideal supply chain network simulator have to achieve good teaching outcomes? (For example: data visualization, team collaboration, competitive modes, customizable scenarios, random events?)
\end{itemize}

\textbf{PART IV: Ideal Simulation System for Teaching Support}
\begin{itemize}
    \item Do you think \ac{llm} can benefit the supply chain simulation? What is your attitudes toward a simulation system where the \ac{llm} models play the role of business managers? Do you think it could perform better than conventional rule-based models or parametric functions that control the company behaviors?
\end{itemize}
  
\begin{itemize}
    \item Apart from the aforementioned features, what other features/functions would make a supply chain simulation most valuable for learning?
\end{itemize}

\subsection{Interview Questions to Students}
\textbf{PART I: Background and Experience}
\textbf{Educational Background}
\begin{itemize}
    \item What industry do you work in? Does your job involve supply chain management?
    \item When did you complete your business program?
    \item Which specific supply chain management courses did you take during your studies?
\end{itemize}
\begin{itemize}
    \item Have you used any supply chain simulation tools during your studies? 
    \item Please provide the system name and describe what it looks like and what it can do?
    \item How did these simulations facilitate learning?
    \item What frustrated you most when using these simulation tools?
\end{itemize}

\textbf{PART II: Evaluation of the Existing Beer Game System}
Here is a standard supply chain simulation used to illustrate the bullwhip effect. The user can play the role of the distributor in a 4-tier supply chain. You are free to explore it.

\begin{itemize}
    \item How intuitive was the interface and navigation?
\end{itemize}
\begin{itemize}
    \item Do you think that this simulation effectively demonstrated the bullwhip effect? Please provide reasons based your experience.
    \item Do you think the system is oversimplified? For example, the system represents supply chain as a linear chain structure instead of a graph structure, making a disconnect from reality. 
    \item What are the potential consequences in terms of educational value if the system is oversimplified?
    \item Is the system capable of handling your learning needs? If yes, can you name some scenarios in workplace that is related to the simulation game.
    \item What concepts do you think would be better demonstrated through a graph-based complex supply chain network?
\end{itemize}

\textbf{PART III: Ideal Simulation System} 
\textbf{Learning Support}
\begin{itemize}
    \item Do you think it valuable to have two map views to display relation structure and the geographic location? How would it benefit the learning experience.
    \item Do you think it valuable to have time reverse functions to check the supply chain status in the past time stamps and enable "what if" analysis by changing he actions in the specific time point.
    \item Do you think it valuable to have the flexibility to modify supply chain networks by adding new companies or shutting down existing companies.
    \item Do you think it is valuable to introduce events into the environment?
    \item Do you think it is valuable to have high flexibility to modify the configuration of companies in the simulated environment (\eg, price, cost, contracts, location, etc.)
    \item Do you expect the simulation system to be capable of serving diverse learning objectives/scenarios? If so, how would you like the system to support it?
    \item Do you expect the simulation system to have strong interactivity with the users? What kind of interaction do you expect to have?
    \item How would you prefer to receive feedback on your decisions?
\end{itemize}

\textbf{Features and Functionality}
\begin{itemize}
    \item Apart from the above-mentioned features, what other features/functions would make a supply chain simulation most valuable for learning?
    \item What other aspects of supply chain simulation tools should we consider?
\end{itemize}

\section{Details of Environment Setup}
\label{appendix:Appendix B}
\subsection{Attribute Variables}
The key variables associated with each agent (\ie, business entity) are defined in the system. These variables describe the state, decisions, and interactions of agents at different stages of the supply chain.

Let $A_{m, i}$ represent the $i^{th}$ agent at stage $m$. The agents in the system are defined by the following variables.

\begin{enumerate}
    \item The supplier set $S_{m,i}$ represents the suppliers of agent $A_{m,i}$. Manufacturers, as the top tier of the supply chain, are assumed to have a fixed source of raw materials for production. Therefore, their supplier sets are denoted as $\mathcal{S}_{A_{M-1,i}}=\emptyset$ in the simulation.
    \item Customer set $D_{m,i}$ includes all the customers of agent $A_{m,i}$. Retailers, as the bottom tier of the supply chain, are assumed to have a fixed group of customers at the local market. Thus, their customer sets are denoted as $\mathcal{D}_{A_{0,i}}={\text{Customer}_i}$ in the simulation.
    \item Inventory level is denoted as 
    \[I=\{I_{m,i}\mid m\in\{0, \ldots, M-1\},\space i\in\{0, \ldots,N_m\}\},\]
    where $I_{m,i}$ represents the quantity of materials held by agent $A_{m,i}$. It follows
    \begin{align}
        I \sim \text{Uniform}(lb_I, ub_I), \quad I_i \in \mathbb{Z}
    \end{align}
    where $lb_I=20$ and $ub_I=25$.
    \item Production capacity is defined as 
    \[C = \{C_{m,i} \mid m\in\{0, \ldots, M-1\},\space i\in\{0, \ldots,N_m\}\},\]
    where $C_{m,i}$ represents the quantity of materials that can be transformed into outgoing products by agent $A_{m,i}$ within a single period. It follows:
    \begin{align}
        C \sim \text{Uniform}(lb_C, ub_C), \quad C_i \in \mathbb{Z}
    \end{align}
    where $lb_C=20$ and $ub_C=40$.
    \item Profit rate $Pr_{m,i}$ is defined as 
    \[Pr = \{Pr_{m,i} \mid m\in\{0, \ldots, M-1\},\space i\in\{0, \ldots,N_m\}\},\]
    where $Pr_{m, i}$ represents the profit rate of agent $A_{m,i}$ for selling one product unit. It follows:
    \begin{align}
        Pr \sim \text{Uniform}(lb_{Pr}, ub_{Pr}), \quad P \in \mathbb{Z}.
    \end{align}
    where $lb_{Pr}=1$ and $ub_{Pr}=2$.
    \item Sale price $P$ is defined as 
    \[P = \{P_{m,i} \mid m\in\{0, \ldots, M-1\},\space i\in\{0, \ldots,N_m\}\},\]
    where $P_i$ represents the price of a unit product that agent $i$ offers to its downstream customers. It is determined by the order cost, production cost and the profit rate as follows:
    \begin{align}
        P = (PC_{m, i} + \max_{j\in {S_{m,i}}}{OC_{i\rightarrow j}^{m}})\cdot Pr_{i}^{m} \quad\forall m\in 0,\ldots, M-2.
    \end{align}
    \item Backlog cost $B$ is defined as 
    \[B = \{b_{m,i} \mid m \in \{0, \ldots, M-1\}, \, i \in \{0, \ldots, N_m\}\},\]
    where $b_{m,i}$ denotes the cost incurred by agent $A_{m,i}$ for each unit of product that has not yet been fulfilled to the downstream customers for one time period. It follows:
    \begin{align}
        B \sim \text{Uniform}(lb_B, ub_B), \quad B \in \mathbb{Z}.
    \end{align}
    where $lb_B=1$, $ub_B=3$.
    \item Holding cost $H$ is defined as 
    \[H = \{h_{m,i} \mid m \in \{0, \ldots, M-1\}, \, i \in \{0, \ldots, N_m\}\},\]
    where $h_{m,i}$ denotes the cost incurred by agent $A_{m,i}$ for storing each unit of product in inventory for one time period. It follows:
    \begin{align}
        H \sim \text{Uniform}(lb_H, ub_H), \quad H \in \mathbb{Z}.
    \end{align}
    where $lb_H=1$ and $ub_H=3$.
    
\end{enumerate}

\subsection{Relational Variables}
The key variables associated with the interactions or
relationships between agents are defined in the system. The generation details are described below.  
\begin{enumerate}
    \item Order cost $OC$ is defined as 
    \begin{equation}
        \begin{aligned}
             OC = \{OC_{A_{m, i}\rightarrow A_{m+1, j}} \mid m &\in\{0, \ldots, M-1\}, \\
        i &\in\{0, \ldots, N_m\}, \\ 
        j &\in\{0, \ldots, N_{m+1}\}\},
        \end{aligned}
    \end{equation}
    where $OC_{A_{m, i}\rightarrow A_{m+1, j}}$ represents the cost paid by agent $A_{m, i}$ to purchase each unit of feed materials from upstream agent $A_{m+1, j}$. It is equal to the sale price offered by upstream supplier $A_{m+1, j}$. I assume the order cost to be 0 (i.e., $OC_{A_{M-1, i}}=0$) for all manufacturers $A_{M-1, i}$.
    \item Lead time is defined as 
    \begin{equation}
        \begin{aligned}
            L = \{L_{A_{m, i}\rightarrow A_{m+1, j}} \mid m &\in\{0, \ldots, M-1\}, \\
            i &\in\{0, \ldots, N_m\}, \\ 
            j &\in\{0, \ldots, N_{m+1}\}\},
        \end{aligned}
    \end{equation}
    where $L_{A_{m, i}\rightarrow A_{m+1, j}}$ represents the number of periods required to deliver products from agent $A_{m, i}$ to agent $A_{m+1, j}$. $L_{A_{m, i}\rightarrow A_{m+1, j}}$ is determined by the distance between two agents as follows:
    \begin{align}
        L_{ij} = \lfloor \sqrt{(x_{m,i} - x_{m+1,j})^2 + (y_{m,i} - y_{m+1,j})^2} \rfloor
    \end{align}
    Here, $(x_i, y_i)$ and $(x_j, y_j)$ represent the locations of agent $A_{m, i}$ and agent $A_{m+1, j}$, respectively. $x_i$, $y_i$, $x_j$, $y_j$ follow \text{Uniform(0, 10)}.
\end{enumerate}

\subsection{Update Rules}
At the end of each period, the variables in the simulated environment are updated to reflect the outcomes of the agents’ decisions. Several rules are defined to govern these updates, which serve to maintain the consistency, track the agents’ states, and propagate the effects of their decisions throughout the environment.

\begin{enumerate}
    \item The actual order fulfillment $R_{A_{m+1,j}\rightarrow A_{m,i}}^t$ is constrained jointly by the upstream supplier $A_{m+1,j}$'s production capacity $C_{m+1,j}$, the inventory level $I_{m+1,j}$, the backlog $B_{m+1,j}$, and the order quantity $O_{m,t}$ placed by the agent $A_{m,i}$ itself. It can be formulated as:
    \begin{equation}
        \begin{aligned}
            R_{m,i}^{t} = \sum_{A_{m+1,j} \in S_{m,i}} \min( 
            & B_{m+1, j}^{t-1} + O_{m,i}^t, \\
            & C_{m+1, j}, \\
            & I_{m+1, j}^{t-1} + R_{m, i}^{t-L_{A_{m+1,j}\rightarrow A_{m, i}}} 
        ),
        \end{aligned}
    \end{equation}
    where $m\in\{0,\ldots,M-2\}$. \\
    Supply of raw materials to the manufacturers are assumed to be unlimited, thus have:
    \begin{align}
        R_{M-1, i}^{t} = O_{M-1, i}^{t}
    \end{align}
    \item The inventory level $I_{m,i}^{t}$ is updated to account for the deliveries received from the suppliers after the lead time $L_{A_{m+1,j}\rightarrow A_{m, i}}$ and the sales during the period $t$ as follows:
    \begin{align}
        I_{m,i}^{t} = I_{m,i}^{t-1} + R_{m, i}^{t-L_{A_{m+1,j}\rightarrow A_{m, i}}} - S_{m,i}^{t},
    \end{align}
    where $m\in\{0, \ldots, M-1\}$.
    \item The sales $S_{m,i}^t$ is determined by the order quantity it can fulfilled to the downstream customers.
    \begin{align}
        S_{m,i}^{t} &= R_{m-1, j}^{t}
    \end{align}
    where $m\in\{1, \ldots, M - 1\}, \quad A_{m-1, j}\in\mathcal{D}_{m,i}^{t}$.\\
    For retailers, their sales $S_{0,i}^t$ is determined by the customer demand as
    \begin{equation}
        \begin{aligned}
            S_{0, i}^{t} = \min(&B_{0, i,}^{t-1} + D_{i}^t,\\
            & C_{0, i}, \\
            & I_{0, i}^{t-1} + R_{0, i}^{t - L_{A_{1,j}\rightarrow A_{0,i}}}), 
        \end{aligned}
    \end{equation}
    where $A_{1,j}\in\mathcal{S}_{0,i}^{t}$.
    \item The backlog $B_{m,i}^t$ at period $t$ depends on the previous backlog, the order $O_{m,i}^{t}$ placed by downstream customers during the current period $t$, and the actual sales $S_{m,i}^t$. It can be formulated as:
    \begin{align}
        B_{m,i}^{t} = B_{m,i}^{t-1} + O_{m,i}^{t} - S_{m,i}^{t}, \quad m\in\{1, \ldots, M-1\},
    \end{align}
    For retailers (\ie, $m=0$), the backlog depends on the customer demand $D_i(t)$ instead of downstream orders. The backlog is express as:
    \begin{align}
        B_{0,i}^{t} = B_{0,i}^{t-1} + D_i(t) - S_{0,i}^{t}
    \end{align}
    \item The profit $Pf_{m,i}$ is a combination of factors. It depends on the sale at the current period $t$ and the price. Additionally, it depends on the order costs incurred to purchase materials from the suppliers during the same period. Furthermore, the inventory costs and backlog costs also contribute to the profit calculation. Overall, it is formulated as:
    \begin{equation}
        \begin{aligned}
            Pf_{i,m}^t &= P_{m,i}\cdot S_{m,i}^{t} \\
            & - \sum_{A_{m+1,j}\in\mathcal{S}_{m,i }^{t}} OC_{A_{m,i}\rightarrow A_{m+1,j}}\cdot R_{m,i}^{t} \\
            & - BC_{m,i}\cdot B_{m,i}^{t}\\
            & - HC_{m,i}\cdot I_{m,i}^{t}, 
        \end{aligned}
    \end{equation}
    where $m\in \{0,\ldots, M-1\}$.
    
\end{enumerate}
\section{Semi-Structured Interview Questions for the User Study}
\label{appendix:Appendix C}  

\textbf{Part 1: Engagement}
\begin{itemize}
    \item The Timeline Tree Map and its color encoding at each state motivates me to improve my decisions for achieving better consequences.
    \item The tiered level of supply chain challenges and the supply chain customization make the simulation feel dynamic and challenging.
    \item I would like to use and create events (e.g., storms, strikes) in the simulation to make it fun and realistic.
    \item The background story and the narratives of the different companies made the experience more immersive.
    \item The advice and encouragement from my AI co-manager (LLM agent) helped me stay motivated, especially when I was struggling.
    \item What was the most engaging or fun feature of the system for you?
    \item Can you describe a moment when you felt particularly motivated or, conversely, demotivated? What caused it?
\end{itemize}

\textbf{Part 2: Learning Support}

\begin{itemize}
    \item The `Performance Quadrant' (on the Operations tab) was an intuitive way to visualize the trade-off between my Fill Rate and Inventory Turns.
    \item The Timeline Tree Map was very helpful for exploring and comparing the outcomes of different decisions by doing time-reverse.
    \item The real-time feedback on profit and loss (\eg, visual encoding on entity icon, the data in analytics dashboard) helps me monitor my performance and motivates me to improve my decisions.
    \item The Analytics Dashboard (Financials) effectively helped me analyze trends and understand the performance of my supply chain regarding inventory and cost management.
    \item The Analytics Dashboard (Operations) effectively helped me analyze trends and understand the performance of my supply chain regarding inventory and order fulfillment.
    \item The Analytics Dashboard (Logistics) effectively helped me analyze trends and understand the performance of my supply chain regarding logistics and supplier reliability.
    \item The reasoning bubbles showing why AI agents made certain decisions were useful for understanding their strategies.
    \item The advice from the LLM co-manager helped me think more critically about my decisions.
    \item Which feature was most helpful for you to analyze the situation and plan your strategy? Why?
    \item Was there any information you needed for your analysis that was missing or hard to access?
    \item Using this system improved my understanding of fundamental supply chain dynamics (\eg, cause-and-effect relationships). 
    \item I have a better understanding of key supply chain metrics (like inventory, backlog, profit) and how they are related after using the simulation. 
    \item The system effectively helped me connect textbook theories with practical decision-making scenarios.
    \item By looking at the charts, I have a much better understanding of the trade-off between holding costs and the risk of stock-outs.
    \item By playing the intermediate level and advanced level, i have a more clear understanding of supply chain resilience and a competing market.
    \item Please describe one key insight or lesson you learned about supply chain management while using SupplyNet.
    \item How does the learning experience with SupplyNet compare to other methods you have used (\eg, lectures, case studies, other simulation games)?
\end{itemize}

\textbf{SUS}
\begin{itemize}
    \item I think I would use this interface frequently.
    \item I found the interface unnecessarily complex. 
    \item I thought the interface was easy to use.
    \item I would need help from a technical person to use this interface.
    \item I found the various functions in the interface well-integrated.
    \item I thought there was too much inconsistency in the interface.
    \item I would imagine most people would learn to use this interface quickly.
    \item I found the interface very cumbersome to use.
    \item I felt confident using the interface.
    \item I needed to learn a lot of things before I could get going with this interface. 
\end{itemize}

\section{Mathematical Formulas for Performance Metrics}
\label{appendix:Appendix D}

This appendix provides the detailed mathematical formulas used in the Task-Oriented Analysis Console to calculate financial, operational, and logistics performance metrics.

\subsection{Financial Performance Metrics}
The financial metrics for an agent $A_{m,i}$ at time $t$ are calculated as follows:
\begin{itemize}
    \item \textbf{Total Profits ($Pf_{m,i}^t$):} The net gain after accounting for sales revenue and all operational costs:
    \[Pf_{m,i}^t = \left( \sum_{A_{m-1,j}\in D_{m,i}} P_{m,i} \cdot S_{m,i}^t \right) - \text{Total Cost}_{m,i}^t\]
    \item \textbf{Total Sales ($\text{Sales}_{m,i}^t$):} The total quantity of products ordered by all downstream customers $D_{m,i}$ at time $t$:
    \[\text{Sales}_{m,i}^t = \sum_{A_{m-1,j}\in D_{m,i}} O_{m-1,j}^t\]
    \item \textbf{Total Cost ($\text{Total Cost}_{m,i}^t$):} The sum of ordering, production, holding, and backlog costs:
    \[\text{Total Cost}_{m,i}^t = \text{OrderCost}_{m,i}^t + \text{ProdCost}_{m,i}^t + \text{HoldCost}_{m,i}^t + \text{BacklogCost}_{m,i}^t\]
    \item \textbf{Ordering Cost ($\text{OrderCost}_{m,i}^t$):} The cost of purchasing materials from upstream suppliers $S_{m,i}$, where $OC_{A_{m,i}\rightarrow A_{m+1,j}}$ is the unit cost and $O_{m,i}^t$ is the order quantity:
    \[\text{OrderCost}_{m,i}^t = \sum_{A_{m+1,j}\in S_{m,i}} OC_{A_{m,i}\rightarrow A_{m+1,j}} \cdot O_{m,i}^t\]
    \item \textbf{Production Cost ($\text{ProdCost}_{m,i}^t$):} Calculated based on the unit production cost $PC_{m,i}$ and the production capacity $C_{m,i}$:
    \[\text{ProdCost}_{m,i}^t = PC_{m,i} \cdot C_{m,i}\]
    \item \textbf{Holding Cost ($\text{HoldCost}_{m,i}^t$):} The cost of storing inventory $I_{m,i}^t$ with unit holding cost $h_{m,i}$:
    \[\text{HoldCost}_{m,i}^t = h_{m,i} \cdot I_{m,i}^t\]
    \item \textbf{Backlog Cost ($\text{BacklogCost}_{m,i}^t$):} The penalty for unfulfilled orders $B_{m,i}^t$ with unit backlog cost $b_{m,i}$:
    \[\text{BacklogCost}_{m,i}^t = b_{m,i} \cdot B_{m,i}^t\]
\end{itemize}

\subsection{Operations Performance Metrics}
The operations performance metrics are calculated as averages over a selected time window $[t_0, t]$ within the timeline tree map:
\begin{itemize}
    \item \textbf{Avg Inventory On Hand ($\overline{I}_{m,i}$):} The average inventory level maintained during the period, calculated based on the selected decision path in the timeline tree map:
    \[\overline{I}_{m,i} = \frac{1}{t - t_0 + 1} \sum_{\tau=t_0}^{t} I_{m,i}^\tau\]
    \item \textbf{Avg Upstream Fill Rate ($\overline{FR}_{m,i}$):} The average ratio of received materials $R_{m,i}^\tau$ to the total quantity ordered $O_{m,i}^\tau$ from all suppliers $S_{m,i}$:
    \[\overline{FR}_{m,i} = \frac{1}{t - t_0 + 1} \sum_{\tau=t_0}^{t} \left( \frac{R_{m,i}^\tau}{O_{m,i}^\tau} \right)\]
    \item \textbf{Avg Upstream On-Time Delivery ($\overline{OTD}_{S}$):} The percentage of deliveries from suppliers that arrived within the expected lead time $L_{A_{m+1,j}\rightarrow A_{m,i}}$. Let $\hat{L}_{A_{m+1,j}\rightarrow A_{m,i}}^k$ be the actual delivery time for the $k$-th order:
    \[\overline{OTD}_{S} = \frac{\text{Count}(\hat{L}^k \leq L)}{\text{Total Deliveries}} \times 100\%\]
    \item \textbf{Avg Downstream On-Time Delivery ($\overline{OTD}_{D}$):} The percentage of the target agent's own deliveries to downstream customers $D_{m,i}$ that were fulfilled within the expected lead time $L_{A_{m,i}\rightarrow A_{m-1,j}}$:
    \[\overline{OTD}_{D} = \frac{\text{Count}(\hat{L}^k \leq L)}{\text{Total Deliveries}} \times 100\%\]
\end{itemize}

\subsection{Logistics Performance Metrics}
The logistics performance metrics provide a high-level overview of the supply chain's efficiency and responsiveness over the period $[t_0, t]$:
\begin{itemize}
    \item \textbf{Avg Fill Rate ($FR_{m,i}$):} The average ratio of fulfilled sales $S_{m,i}^\tau$ to the total order quantity $O_{m,i}^\tau$ received from downstream customers:
    \[FR_{m,i} = \frac{1}{t - t_0 + 1} \sum_{\tau=t_0}^{t} \frac{S_{m,i}^\tau}{O_{m,i}^\tau}\]
    \item \textbf{Average Inventory ($\overline{I}_{m,i}$):} The average quantity of products held in stock over the time period:
    \[\overline{I}_{m,i} = \frac{1}{t - t_0 + 1} \sum_{\tau=t_0}^{t} I_{m,i}^\tau\]
    \item \textbf{Average Inventory Turns ($IT_{m,i}$):} A standard measure of how many times the inventory is sold and replaced over the period. It is calculated as the total Cost of Goods Sold (COGS) divided by the average inventory value. The unit cost is defined as the sum of unit ordering cost, unit production cost, and unit holding cost ($PC_{m,i} + \max_{j\in S_{m,i}} OC_{A_{m,i}\rightarrow A_{m+1,j}} + h_{m,i}$):
    \[IT_{m,i} = \frac{\sum_{\tau=t_0}^{t} \text{Total Cost}_{m,i}^\tau}{\overline{I}_{m,i} \cdot (PC_{m,i} + \max_{j\in S_{m,i}} OC_{A_{m,i}\rightarrow A_{m+1,j}} + h_{m,i})}\]
    \item \textbf{Average Weeks of Supply ($WOS_{m,i}$):} A standard measure of how many weeks (or time periods) the current inventory will last based on average demand:
    \[WOS_{m,i} = \frac{\overline{I}_{m,i}}{\left( \sum_{\tau=t_0}^{t} S_{m,i}^\tau \right) / (t - t_0 + 1)}\]
\end{itemize}

\section{LLM Prompt Infrastructure for Decision Making}
\label{appendix:Appendix E}

This appendix details the prompt construction and LLM integration within the SupplyNet framework. It clarifies the role of the "Golden Rules," provides examples of prompt templates used for the textualization of extracted subgraphs, and demonstrates how the exact natural language outputs from the LLM are converted into numerical simulation decision variables.

\subsection{Content and Role of the "Golden Rules"}
The "Golden Rules" serve as foundational operational principles that equip the LLM agents with essential domain knowledge in supply chain management. By embedding these rules into the prompt, we ensure that the agents' decision-making processes are grounded in established economic and operational logic, such as inventory equilibrium, profitability margins, and the mitigation of the bullwhip effect.

\begin{lstlisting}[language=Python]
GOLDEN_RULES = (
    "Operational Principles for Inventory Optimization:\n"
    "1. Inventory Equilibrium: Open Orders must equal 'Expected Downstream Demand + Backlog' to ensure stability.\n"
    "2. Profitability Margin: Sales Price must exceed the aggregate of Production and Procurement Costs.\n"
    "3. Procurement Constraints: Order placement is restricted to identified upstream suppliers only.\n"
    "4. Lead-Time Mitigation: Proactive order placement is required to account for supplier lead times.\n"
    "5. Strategic Sourcing: Optimize supplier selection based on lead-time efficiency vs. procurement costs.\n"
    "6. Bullwhip Effect Suppression: Distribute order volumes across multiple periods to minimize variance amplification."
)
\end{lstlisting}
\revision{
\subsection{Textualization of Extracted Subgraphs}
To provide the LLM with the current supply chain status, we extract the relevant subgraph for each agent and convert it into a structured natural language format (Textualization). This graph-based environment status is represented as two markdown tables: a Node Table for agent attributes (e.g., inventory, costs) and an Edge Table for relational properties (e.g., lead times, delivery status).
}
\begin{lstlisting}[language=Python]
NODE_TABLE_TEMPLATE = """
### Node Table (Agent Attributes)
| Node ID | Attributes (<node label>, <role>, <inventory level>, <sale price>, <costs>) |
|:--------|:--------------------------------------------------------------------------|
| <id>    | Inventory: <value>, Price: <value>, Production Cost: <value>, Order Cost: <value> |
"""

EDGE_TABLE_TEMPLATE = """
### Edge Table (Supply Chain Relations)
| Source  | Destination | Attributes (<lead time>, <order placement>, <delivery status>) |
|:--------|:------------|:--------------------------------------------------------------|
| <id>    | <id>        | Lead Time: <value> |
| <id>    | <id>        | On-the-way delivery: <value> |
"""
\end{lstlisting}
\revision{
\subsection{LLM Output and Conversion to Simulation Variables}
To ensure that the natural language generated by the LLM can be seamlessly integrated back into the simulation engine, we enforce a strict JSON output schema. The LLM is instructed to output exact numerical parameters (e.g., integer values for order quantities) and specific string identifiers (e.g., supplier IDs for network configuration). The simulation engine directly parses this JSON object to update the environment state for the next timestep.
}
\begin{lstlisting}[language=Python]
JSON_OUTPUT_FORMAT = {
    "supplier_selection": {
        "reasoning": "<1-2 sentence justification>",
        "removals": ["<id>"],
        "additions": ["<id>"]
    },
    "order_placement": {
        "orders": [
            {"supplier_id": "<id>", "quantity": "<int>"}
        ]
    },
    "performance_review": {
        "business_performance": "<string_analysis_max_15_words>",
        "feeling": "<satisfied/concerned>"
    }
}
\end{lstlisting}
\revision{
\subsection{Integrated Prompt Example}
The final prompt assembled for the LLM integrates the system role, the textualized graph data, contextual updates (events and downstream demands), the Golden Rules, and the strict output instructions.
}
\begin{lstlisting}[language=Python]
TASK_DESCRIPTION = (
    "Task I: Supply Network Configuration - Evaluate upstream edges and modify the supplier list.\n"
    "Task II: Procurement Execution - Determine optimal order quantities and assess round performance."
)

prompt = f"""
### SYSTEM ROLE
You are the lead Supply Chain Manager for {agent_role}. 
Simulation Round: {round_period}.

### ENVIRONMENT STATUS (GRAPH REPRESENTATION)
{node_data}
{edge_data}

### CONTEXTUAL UPDATES
- Emergent Events: {event_msg}
- Downstream Requirements: {downstream_msg}

---
### OPERATIONAL CONSTRAINTS
{GOLDEN_RULES}
---

### DECISION TASKS
{TASK_DESCRIPTION}

### OUTPUT INSTRUCTIONS
Return your decisions strictly in the following JSON format. Ensure all IDs match the provided tables.
{JSON_OUTPUT_FORMAT}
"""

\end{lstlisting}
\ques{[R1-Q3, R1-Q4]}
\section{Implementation Details}
\label{appendix:Appendix F}

The prototype of SupplyNet is implemented using a modern web technology stack, ensuring a responsive and interactive user experience.

\textbf{Frontend and Visualization:} The user interface is built with Vue 3.5 and Vite 6.1, utilizing Element Plus for UI components. The interactive supply chain network is rendered using Vue Flow (\texttt{@vue-flow/core}), which supports drag-and-drop interactions and dynamic node updates. Data visualization in the Analysis Console is powered by Chart.js and ECharts, providing real-time interactive charting capabilities. The onboarding tutorial is implemented using Driver.js, and internationalization is supported via \texttt{vue-i18n}.

\textbf{Backend and Simulation Core:} The backend simulation engine is implemented in Python (version 3.14+). It manages the global state of the supply chain, calculates performance metrics based on standard formulas, and orchestrates the timestep progression. Communication between the frontend and the backend is handled via Axios over RESTful APIs.

\textbf{LLM Agents:} We developed a custom multi-agent framework tailored specifically for this supply chain simulation. All agents in the system are powered by OpenAI's \texttt{gpt-4o} model to ensure consistent and high-quality reasoning capabilities across the simulation. The agents receive contextualized subgraphs and task descriptions, and return structured JSON responses that are directly parsed by the simulation core.

\textbf{Database and State Management:} The supply chain state for each timestep is persisted as JSON files. These files store comprehensive agent properties, including costs, profits, inventory levels, and supply relationships. This lightweight approach allows for efficient state retrieval and supports the branching mechanism in the Timeline Tree Map.

\textbf{System Requirements and Token Costs:} Running the system locally requires Node.js v22 and Python 3.14+, with a minimum of 8GB RAM recommended. Since the LLM reasoning is performed via cloud APIs, no dedicated GPU is required. An active OpenAI API Key is necessary to run the simulation. Based on current pricing, a typical simulation session processing 100k to 200k tokens costs approximately \$0.25 to \$2.00. While the system can be configured to use smaller local LLMs to eliminate costs, the quality of complex supply chain reasoning and simulation stability may be degraded compared to \texttt{gpt-4o}.
\ques{[R2-W3]}





\end{document}